\documentstyle[12pt]{article}
\input{amssym.def}
\input{amssym}
\def\be{\begin{equation}}
\def\ee{\end{equation}}

\def\lb{\label}
\def\ba{\begin{eqnarray}}
\def\ea{\end{eqnarray}}
\def\D{{\cal D}}

\def\theequation{\thesection.\arabic{equation}}

\begin{document}

\vspace{2cm}

\begin{center}
\Large{ DIFFUSION ALGEBRAS}
\end{center}

\vspace{.5cm}

\begin{center}
\large{A.P.\,ISAEV${}^{a \, 1}$, P.N.\,PYATOV${}^{a,b \, 2}$
and  V.\,RITTENBERG${}^{c \,3}$ }
\end{center}

\begin{center}
${}^{a}$ Bogoliubov Laboratory of Theoretical Physics,
Joint Institute for Nuclear Research,
Dubna, Moscow region 141980, Russia. \\[2mm]
${}^{b}$ Max-Planck-Institut f\"{u}r Mathematik,
Vivatsgasse 7, 53111 Bonn, Germany. \\[2mm]
${}^{c}$
Physikalisches Institut, Universit\"{a}t Bonn,
Nussallee 12, 53115 Bonn,
Germany.
\end{center}

\bigskip
\vspace{1cm}

\begin{center}
ABSTRACT
\end{center}

We define the notion of "diffusion algebras". They are quadratic
Poincar\'e-Birkhoff-Witt (PBW)
algebras which are useful in order to find exact expressions for the
probability distributions of stationary states appearing in
one-dimensional stochastic processes with exclusion. One considers processes
in which one has N species, the number of particles of each
species being conserved.
 All diffusion algebras are obtained. The known
examples already used in applications are special cases in our
classification. To help the reader interested in physical problems,
the cases $N=3$ and $4$ are listed separately.

\vspace{5cm}
${}^1$ E-mail: isaevap@thsun1.jinr.ru

\vspace{1cm}
${}^2$ E-mail: pyatov@thsun1.jinr.ru; pyatov@mpim-bonn.mpg.de

\vspace{1cm}
${}^3$ E-mail: vladimir@th.physik.uni-bonn.de

\pagebreak

\section{Introduction.}
\setcounter{equation}0

One-dimensional stochastic processes with random-sequential updating
have stationary states (far away form
equilibrium) with very interesting physical properties. One observes phase
transitions about which still little is known. As opposed to equilibrium
states where the phase transitions are essentially a bulk phenomenon, in
stationary states, the boundary conditions play an essential role \cite{D}.
One observes bulk induced phase transitions if one looks to problems on a
ring \cite{E},
boundary induced phase transitions in the case of open systems \cite{SD},
or a combination of both \cite{ARR}. When dealing with phase transitions, one is
interested in exact expressions for the relevant physical quantities like
the current densities (which are zero for equilibrium problems) and
various correlation functions. A major step in this direction was achieved
when it was understood that in certain cases one can use the so-called
"matrix product states" approach \cite{DEHP,D}. In our opinion, from a mathematical
point of view, this approach was loosely defined
(for an illustration see Appendix A). The aim of this paper is
to improve upon this situation and, as a bonus, to give a large number of
processes where one can find "matrix product states". We define
a class of quadratic algebras to which we have coined the
name "diffusion algebras". We also show how to construct all of them.
We believe that
in this way one can
bring some mathematical "beauty" in what
was not, up to now, a systematic approach.

We start with the physical problem in order to explain the motivation of our mathematical work.

In the present paper we consider a restricted class of processes
in which we take $N$ species of particles with $N-1$ conservation
laws in the bulk. We take a one-dimensional lattice with $L$ sites
and assume that one has only nearest-neighbor interaction with
exclusion (there can be only one particle on a given site). In the time interval $dt$ only the following processes
are allowed in the bulk:
\be
\lb{1.1}
\alpha + \beta \rightarrow
\beta + \alpha \;\;\; (\alpha,\beta = 0,1, \dots , N-1) \; ,
\ee
with the probability $g_{\alpha \beta} \, dt$ ($g_{\alpha\beta}\geq 0$).
Eq.(\ref{1.1}) is a symbolic equation used by physicists. Its meaning is that the particles $\alpha$ and $\beta$ on successive sites exchange their places.
Obviously the number of
particles $n_\alpha$ of each species $\alpha$ are conserved
$(\sum_{\alpha =0}^{N-1} \, n_\alpha = L)$. One is interested in
the probability distribution $P(\alpha_1 , \alpha_2 , \dots ,
\alpha_L)$, $(\alpha_k = 0,1, \dots , N-1)$
 for the stationary state. In order to obtain it, in
the matrix product approach, one considers $N$ matrices
$\D_\alpha$ and $N$ matrices
$X_\alpha$
acting in an auxiliary vector space and
satisfying the following relations \cite{S}:
\be
\lb{1.2}
\begin{array}{c}
g_{\alpha \beta} \D_{\alpha} \D_{\beta} -
g_{\beta \alpha} \D_{\beta} \D_{\alpha} = \frac{1}{2} \,
\{ X_\beta , \, \D_\alpha \} - \frac{1}{2} \,
\{ X_\alpha , \, \D_\beta \} \; ,  \\ \\
\left[\D_\alpha , \, X_\beta \right] = -
\left[ \D_\beta , \, X_\alpha \right] \; .
\end{array}
\ee
In Eqs.(\ref{1.2}) $\{. ,. \}$ represent
anti-commutators.

If one considers processes on a ring (periodic
boundary condition), the un-normalized probability distribution
has the following expression \cite{DJLS}:
\be
\lb{1.3}
P(\alpha_1 , \alpha_2
, \dots , \alpha_L) = Tr \left(\D_{\alpha_1}  \D_{\alpha_2}
\dots  \D_{\alpha_L} \right)
\ee
Notice that the matrices $X_\alpha$
don't appear in Eq.(\ref{1.3}). The expression of the probability
distribution is special in at least two ways. As a consequence of
the conservation laws, Eq.(\ref{1.3}) connects only monomials
with the same numbers
$n_0, n_1, \dots , n_{N-1}$ of generators
$\D_0,\D_1, \dots , \D_{N-1}$. This implies that one can
use different matrices for monomials for different values of
the set $n_0,n_1, \dots , n_{N-1}$.
This observation is important since for example, a given infinite-dimensional
representation can have a finite trace for certain class of monomials but
can diverge for another class of monomials. In order to use Eq.(\ref{1.3}), for
the latter one can use a different representation which for example,
is traceless for the first class of monomials but has a finite trace for
the second class.
This problem will be
explained in detail in \cite{IPR}.

On the other hand if different
representations can be used for all monomials one has the remarkable property that up to a
factor (the expression (\ref{1.3}) does not contain a
normalization factor), the traces are independent on the
representations one uses.
For concrete calculations one takes therefore the representation with
the smallest dimension.

If one considers open systems, the bulk processes have to be completed by
boundary processes (they break the conservation laws). On the left side of
the chain (site $1$) and on the right side of the chain (site $L$) we assume
that in the time interval dt, the particle $\alpha$ is replaced by the particle $\beta$:
\be
\lb{1.4}
\alpha \longrightarrow \beta \ ,
\ee
with the probabilities
\be
\lb{1.5}
L^\alpha_\beta \, d \, t  \ , \quad\mbox{respectively}\quad
R^\alpha_\beta \, d \, t  \ .
\ee
The matrices appearing in (\ref{1.5}) are
intensity matrices \cite{ADR} for which the
diagonal elements are given by the non-diagonal ones:
\be
\lb{1.6}
L^\alpha_\alpha = - \sum_{\beta \neq \alpha} \, L^\alpha_\beta \; , \;\;\;
R^\alpha_\alpha = - \sum_{\beta \neq \alpha} \, R^\alpha_\beta \; ,
\ee
The non-diagonal matrix elements are non-negative.

For open systems,
the un-normalized probability distribution functions are given by a matrix
element in the auxiliary vector space:
\be
\lb{1.7}
P(\alpha_1 , \alpha_2 , \dots , \alpha_L) =
{\cal h} 0 | \, \D_{\alpha_1}  \D_{\alpha_2}  \dots  \D_{\alpha_L} \,
| 0 {\cal i}\ ,
\ee
where the bra (ket) state ${\cal h}0|$ (respectively $|0{\cal i}$) are given by the
following conditions:
\be
\lb{1.8}
{\cal h} 0 | \, ( L^\beta_\alpha \, \D_\beta + X_\alpha ) = 0 \; , \;\;\;
( R^\beta_\alpha \, \D_\beta - X_\alpha ) \, | 0 {\cal i} = 0 \; ,
\ee
The expressions (\ref{1.3}) and (\ref{1.7})
for the probability distributions are of
little use unless one finds matrices which satisfy the conditions
(\ref{1.2}) and
have a trace, or one finds matrices which satisfy (\ref{1.2}) and have the
property (\ref{1.8}). In a very nice paper \cite{KS} it was shown that for
arbitrary bulk and boundary rates one can in principle construct infinite
dimensional matrices which satisfy the conditions
(\ref{1.2}) and (\ref{1.8}) (if they
have also finite traces is unclear).
It is however practically impossible to obtain explicit expressions for these matrices.
It is also not clear which supplementary relations
besides (\ref{1.2}) they satisfy.
In other words, the algebraic structure behind the
relations (\ref{1.2}) is obscure.

In the present paper we adopt a different
approach. We start by searching for
quadratic algebras which are of PBW type
(this notion is explained in the beginning of Section 2)
with a structure which is understood and look for those bulk rates
for which a simplified version of the relations (\ref{1.2}) exist. Then we look
to boundary conditions compatible with the simplified version of
Eq.(\ref{1.8}).  We make the Ansatz:
\be
\lb{1.9}
X_\alpha = x_\alpha \, e \; ,
\ee
where $e$ acts as a unit element on $\D_\alpha$
\be
\lb{1.10}
e \, \D_\alpha = \D_\alpha \, e  = \D_\alpha  \; ,
\ee
and $x_\alpha$ are c-numbers.
With this Ansatz, instead of Eq.(\ref{1.2}) one obtains
$N(N-1)/2$ relations for the matrices $\D_\alpha$:
\be
\lb{algebra}
g_{\alpha \beta} \D_{\alpha} \D_{\beta} -
g_{\beta \alpha} \D_{\beta} \D_{\alpha} =
x_\beta \D_\alpha - x_\alpha \D_\beta \; ,
\ee
and instead of Eq. (\ref{1.8}) one obtains:
\be
\lb{1.8a}
{\cal h} 0 | \, ( L^\beta_\alpha \, \D_\beta + x_\alpha \, e) = 0 \; , \;\;\;
( R^\beta_\alpha \, \D_\beta - x_\alpha \, e) \, | 0 {\cal i} = 0 \; ,
\ee
Taking into account the relations (\ref{1.6}), one has
$$
\sum_{\alpha = 0}^{N-1} \, x_\alpha = 0 \; .
$$

The consequence of the Ansatz (\ref{1.9}) will be that we will not be able to
find expressions for the probability distributions for arbitrary bulk and
boundary rates. Further limitations on the possible bulk rates  will
appear when we ask for the relations (\ref{algebra}) to define quadratic algebras
of PBW type. This implies that the ordered monomials
\be
\lb{1.12}
\D_{N-1}^{n_0} \, \D_{N-2}^{n_1} \cdots \, \D_{0}^{n_{N-1}}
\ee
form a basis in the algebra. Algebras of this type will be called
"diffusion algebras".

Once the algebras are known, we can ask if they are also useful for
applications to stochastic processes. First one has to find for which
algebras one can choose all the $g_{\alpha\beta}$ non-negative. Next, since
the relation (\ref{algebra}) gives recurrence relations among traces of different
monomials, one has to find in which cases  one has not only the trivial
solution for these recurrence relations (all the traces vanish). These
cases can be used to study stochastic processes on a ring. Finally, one
has to find for which boundary matrices one can find representations for
which the conditions (\ref{1.8a}) are satisfied.
This is not a trivial exercise.
For the cases for which one finds solutions, one can get then the
expression of the probability distribution for the open system.

 We would like to give a supplementary argument which motivated us to
look for PBW algebras. Interesting physics appears when one has only
infinite dimensional representations either for the ring problem or for
the open system \cite{D,E,DEHP}. The PBW algebras have at least one
infinite-dimensional representation (the regular one), which doesn't mean
that, in general, they don't have also finite-dimensional ones.

In the present paper we concentrate on the diffusion algebras only. In a
sequel \cite{IPR} we will discuss in more detail the representation theory.
We will also not look for physical applications except for stressing the cases when the rates $g_{\alpha\beta}$ can't be chosen non-negative.

In Sec.2 we will show that in order to have a
diffusion algebra, the $g_{\alpha \beta}$ and
$x_\alpha$ have to satisfy certain identities
which are the equivalent of the Jacobi identities for the structure
constants in the case of Lie algebras
\footnote{Correctly speaking, we mean universal enveloping algebras
of Lie algebras}.
It is remarkable that, as we are going to show in the next sections,
we are able to find all the solutions of these identities.
Inspecting the Eqs.(\ref{algebra}) one
notices that if some of the $x_\alpha$ are not zero they can be rescaled in
the definition of the generators $\D_\alpha$.
The number of non-vanishing $x_\alpha$
will play an important role in the classification of diffusion algebras.

In Sec.3 we consider the case $N=3$ corresponding to the three species problem
(the case $N=2$ is well known \cite{ER}). The case $N=3$ is not only interesting on its
own but is relevant for understanding the case for arbitrary $N$ described
in Sec. 4. We are going to recapture all the known
examples \cite{E,ADR,AHR,K}
which are interesting for applications and get a few new
algebras which are interesting on their own.

We would like to mention that the case $N=3$ was approached previously
from two other points of view. In Ref.\cite{ADR},
one has looked directly to the
open system,  and classified the type of boundary matrices appearing in
Eq.(\ref{1.8a}). This is possible since the boundary matrices are intensity
matrices which, as explained in Ref.\cite{ADR}, have special properties. Next, one
has looked for bulk rates (see Eq.(\ref{algebra})) compatible with the boundary
conditions.
The problem on the ring was considered in
Ref.\cite{AHR}. Here one has asked which relations (\ref{algebra}) are compatible with
the trace operation and one has given the smallest representations (this was enough for applications).

In Appendix A we present an instructive different (although equivalent)
approach to obtain part of the diffusion algebras for $N=3$. We also show a
natural way to define a quotient of one of the algebras.

In Sec.4 we consider the case of $N$ generators. First we give seven series
of algebras and then present a theorem which allows to find
all the diffusion algebras of PBW type. The proof of this theorem would imply a long
discussion and would not fit in this paper which is also
aimed at physicists who are not necessarily interested in
combinatorics. We have therefore decided to publish it separately.
Some algebras are not suitable for
applications to stochastic processes (they are not compatible with
positive rates).

In order to help the reader who is interested in applications and not in
mathematics, in Appendix B we list the diffusion algebras with positive
rates for $N=4$.

In Sec.5 several physically meaningful generalizations of the diffusion
algebras are discussed. A possible connection between diffusion algebras
and quantum Lie algebras is also pointed out.

\section{Diamond conditions.}
\setcounter{equation}0

We shall search for PBW-type associative  algebras which are generated
by the unit $e$ and the elements $\D_{\alpha}$, $\alpha = 0,1,\dots ,N-1$
satisfying  $N(N-1)/2$ quadratic-linear relations
given by (\ref{algebra}).

The PBW property (see, e.g., Sec.3 of the Ref.\cite{Bergman}) implies that
given a set of generators $\{\D_{\alpha}\}$,
one can express any element of the algebra as a linear combination
of ordered monomials in $\D_{\alpha}$.
Furthermore, all the ordered monomials are assumed to be linearly
independent.\footnote{
The term PBW is due to Poincar\'{e}, Birkhoff and Witt \cite{PBW}
who describe a linear basis in a universal enveloping algebra.
}

For concreteness let us fix an alphabetic order
\be
\lb{order}
\D_{\beta} > \D_{\alpha}
\;\;\; {\rm if} \;\;\;
\beta > \alpha \ .
\ee
Then, a linear basis for the PBW-type algebra is given by the unit $e$ and
the set of monomials
\be
\lb{basis}
\D_{\alpha_1}^{n_1} \D_{\alpha_2}^{n_2} \cdot \dots \cdot
\D_{\alpha_k}^{n_k}\ , \quad k=1,2,\dots \ ,
\ee
where $\alpha_1 > \alpha_2 > \dots > \alpha_k$ and
$n_1, n_2, \dots ,n_k$ are arbitrary positive integers.

Imposing the PBW condition for
the diffusion algebra (\ref{algebra})
we first demand
\be
\lb{PBW0}
g_{\alpha\beta}\neq 0, \quad\forall\;\;  \alpha < \beta
\ee
in order to be able to express any polynomial in $\D_{\alpha}$ as
a linear combination of the basic monomials (\ref{basis}).

Next, using (\ref{algebra}) one can reorder any cubic monomial
$\D_{\alpha}\D_{\beta}\D_{\gamma}\rightarrow\D_{\gamma}\D_{\beta}\D_{\alpha}$,
where $\alpha <\beta <\gamma$
in two different ways:

\unitlength=6mm
\begin{picture}(17,4)
\put(1.5,1.5){$\D_\alpha \D_\beta \D_\gamma$}
\put(4.5,2){\vector(1,1){1}}
\put(4.5,1.4){\vector(1,-1){1}}
\put(6,3){$\D_\alpha \D_\gamma \D_\beta$}
\put(6,0){$\D_\beta \D_\alpha \D_\gamma$}
\put(9,3.2){\vector(1,0){1}}
\put(9,0.2){\vector(1,0){1}}
\put(10.5,3){$\D_\gamma \D_\alpha \D_\beta$}
\put(10.5,0){$\D_\beta \D_\gamma \D_\alpha$}
\put(13.5,3){\vector(1,-1){1}}
\put(13.5,0.4){\vector(1,1){1}}
\put(15,1.5){$\D_\gamma \D_\beta \D_\alpha$}
\end{picture}

Demanding the coincidence of the resulting expressions
for $\D_{\alpha}\D_{\beta}\D_{\gamma}$ in terms of ordered monomials
one obtains the relation
\ba
\nonumber
&&
\phantom{a}\hspace{-10mm}
x_\alpha \, g_{\gamma \beta} ( \Lambda_{\alpha \beta \gamma} +
\Lambda_{\gamma \beta} ) \D_\gamma \D_\beta  +
x_\beta \, g_{\gamma \alpha} ( \Lambda_{\alpha \beta \gamma} +
\Lambda_{\alpha \gamma} ) \D_\gamma \D_\alpha
\\
\nonumber
&&+\ x_\gamma \, g_{\beta \alpha} ( \Lambda_{\alpha \beta \gamma} +
\Lambda_{\beta \alpha} ) \D_\beta \D_\alpha
+ x_\alpha x_\beta ( g_{\gamma \alpha}
- g_{\gamma \beta} - \Lambda_{\alpha \beta \gamma} ) \, \D_\gamma
\\
\lb{rel1}
&&
+\ x_\alpha x_\gamma ( g_{\gamma \beta} - g_{\beta \alpha}) \, \D_\beta
+ x_\alpha x_\beta (g_{\beta \alpha}
-g_{\gamma \alpha} + \Lambda_{\alpha \beta \gamma} ) \, \D_\alpha\ =\ 0
\; ,
\ea
which results in the following six conditions for the $g_{\alpha \beta}$'s
and the $x_\alpha$'s
\ba
\lb{PBW1}
x_\alpha \, g_{\gamma \beta} ( \Lambda_{\alpha \gamma} -
\Lambda_{\alpha \beta} ) = 0\ ,
\\
\lb{PBW2}
x_\beta \, g_{\gamma \alpha} ( \Lambda_{\beta \gamma} +
\Lambda_{\alpha \beta} ) = 0\ ,
\\
\lb{PBW3}
x_\gamma \, g_{\beta \alpha} ( \Lambda_{\alpha \gamma} -
\Lambda_{\beta \gamma} ) = 0\ ,
\\
\lb{PBW4} x_\beta x_\gamma \, (
\Lambda_{\beta \gamma} + g_{\alpha \beta } -
g_{\alpha \gamma}  ) = 0\ ,
\\
\lb{PBW5}
x_\alpha x_\gamma \, ( g_{\beta \alpha} - g_{\gamma \beta} ) = 0\ ,
\\
\lb{PBW6}
x_\alpha x_\beta \, ( \Lambda_{\alpha \beta } +
g_{\beta \gamma} - g_{\alpha \gamma}  ) = 0\ , && \forall\; \alpha <\beta
<\gamma\ .
\ea
In Eqs. (\ref{rel1}) and
(\ref{PBW1})-(\ref{PBW6}) we have
introduced the notation
\be
\lb{notation}
\Lambda_{\alpha \beta} := g_{\alpha \beta} - g_{\beta \alpha}\ , \qquad
\Lambda_{\alpha \beta \gamma} := \Lambda_{\alpha \beta}
+ \Lambda_{ \beta \gamma} + \Lambda_{\gamma \alpha}\ .
\ee

One can show that the conditions
(\ref{PBW1})--(\ref{PBW6}) are necessary in order
to avoid linear dependences between ordered quadratic (or even
first order) monomials in $\D_\alpha$, $\D_\beta$ and $\D_\gamma$.
One can improve this result applying the diamond Lemma
\cite{Bergman} to our concrete case. Namely, the algebra
(\ref{algebra}) possesses the PBW property iff the conditions
(\ref{PBW1})--(\ref{PBW6}) are fulfilled. We will refer relations
(\ref{PBW1})--(\ref{PBW6}) as "diamond conditions" in what
follows.

In the next sections we are going to find the solutions of the
diamond conditions first for the case $N=3$
and then in the general case.

\section{Classification of diffusion algebras with $3$ generators.}
\setcounter{equation}0

The classification of diffusion algebras generated by an arbitrary number
$N$ of elements $\D_\alpha$ proceeds as follows.

First, one notes that any  subset of $k<N$ elements
$\D_\alpha$  generates a
subalgebra of (\ref{algebra}) which is again a diffusion algebra.
So, it looks natural to begin with minimal size subalgebras
produced by 3 generators (for the case of
2 generators one does not have any nontrivial diamond conditions to solve).
Then, a closer inspection of the diamond relations (\ref{PBW1})--(\ref{PBW6})
shows that they would be fulfilled for the whole algebra (\ref{algebra})
provided that they are satisfied for all the minimal size subalgebras.

So, we shall start by classifying diffusion algebras
generated by three elements $\D_{\alpha}, \D_{\beta}, \D_{\gamma}$,~ $\alpha <
\beta < \gamma$. It is natural to fix the values of indices as
$\alpha=0$, $\beta=1$, $\gamma=2$ (this choice is adopted in Appendix A).
Here however we will not assign concrete values to
the indices $\alpha$, $\beta$ and $\gamma$
keeping in mind that for general $N > 3$
$\alpha<\beta<\gamma$ may denote any triple of
indices from the set $0,1,\dots ,N-1$.

Depending on how many of the parameters
$x_{\alpha}, x_{\beta}, x_{\gamma}$
take nonzero values
the classification falls into
four cases. Namely, when all three parameters
$x_{\alpha} , x_{\beta}$ and $x_{\gamma}$ are nonzero we obtain the
algebras of type $A$. If one of the $x$'s is zero and the remaining
two are not equal to zero we have the algebras of type $B$. The algebras
with only one nonzero parameter $x$ and those with all $x$'s  zero
are called algebras of type $C$ and respectively $D$.

Now we consider in details the algebras of type $A$ -- $D$.
\vspace{0.2cm}

\noindent
{\bf Case A.} \underline{All $x_{\alpha} , x_{\beta}$ and
$x_{\gamma}$ are nonzero.}
\vspace{0.3cm}

Eqs.(\ref{PBW4})--(\ref{PBW6}) give us two constraints
\be
\lb{a1}
g_{\beta\alpha} = g_{\gamma\beta} = g_{\alpha\beta} + g_{\beta\gamma} -
g_{\alpha\gamma}\ .
\ee
Then, there are two possibilities.

\vspace{0.2cm}
\noindent
{\bf 1).}~
If $g_{\beta\alpha}=g_{\gamma \beta} \neq 0$,
the Eqs.(\ref{PBW1})
and (\ref{PBW3}) give
$\Lambda_{\alpha \gamma} =
\Lambda_{\alpha \beta} = \Lambda_{\beta\gamma}$
whereof one obtains
\be
\lb{a2}
g_{\gamma\alpha} = g_{\beta\gamma} = g_{\alpha\beta}\ .
\ee
In view of (\ref{PBW0}) one has $g_{\gamma\alpha} \neq 0$ and,
therefore, from eq.(\ref{PBW2}) one obtains
\be
\lb{a3}
\Lambda_{\beta\gamma}=-\Lambda_{\alpha\beta}\ .
\ee
Finally, from (\ref{a1}), (\ref{a2}) and (\ref{a3}) one concludes
\be
\lb{a4}
g_{ij} = g \neq 0\ , \quad \forall \; i,j \in \{\alpha,\beta,\gamma\}\ .
\ee
The corresponding $A_{I}$-type algebra is
\be
\lb{a5}
{\bf A}_{\bf I} \;\;\;
\mbox{\framebox[110mm]{$
g \, [ \D_i , \, \D_j ]  = x_j \, \D_i -  x_i \, \D_j\ ,
\quad \forall \; i\neq j\in
\{\alpha,\beta,\gamma\}\ ,\; g\neq 0\ .
$}}
\ee
These are relations of Lie algebraic type. By rescaling
the generators $\D_i \rightarrow {x_i\over g} E_i$
one can remove all the parameters from (\ref{a5}) and obtain:
$$
[ E_i , \, E_j ]  = E_i -  E_j\ .
$$

\vspace{0.2cm}
\noindent
{\bf 2).}~ If $g_{\beta\alpha} = g_{\gamma\beta} = 0$, using Eq.(\ref{a1})
one transforms the only remaining nontrivial equation (\ref{PBW2}) as
$$
g_{\gamma\alpha}(\Lambda_{\beta\gamma}+\Lambda_{\alpha\beta}) =
g_{\gamma\alpha}(g_{\beta\gamma}+g_{\alpha\beta}) = g_{\gamma\alpha}\
g_{\alpha\gamma}=0\qquad \Rightarrow \qquad g_{\gamma\alpha}=0\ .
$$
The corresponding $A_{II}$-type algebra is:
\be
\lb{a6}
{\bf A}_{\bf II} \;\;
\mbox{\framebox[110mm]{$
\begin{array}{rl}
& g_{ij} \D_i \D_j  =
x_j \, \D_i -  x_i \, \D_j\ ,
\quad \forall\; i<j\in \{\alpha,\beta,\gamma\}\ , \\[2mm]
\mbox{where}\;\;&
g_{ij}:= g_i - g_j\ , \quad g_i\neq g_j\;\; \forall\; i\neq j\ .
\end{array}
$}}
\ee
The parameters $g_i$ ($i\in\{\alpha,\beta,\gamma\}$)
introduced in Eq.(\ref{a6}) above are defined up to
a common shift $g_i \rightarrow g_i + c$.

This algebra is invariant under the transformation
$$
\D'_i = \D_i + \frac{x_i}{g_i - y} \; , \;\;\;
g'_i = \frac{1}{y - g_i} \; , \;\;\;
x'_i = \frac{x_i}{(y - g_i)^2} \; ,
$$
where $y$ is an arbitrary parameter.
As explained in Appendix A, the algebra (\ref{a6}) has a natural
quotient $[\D_\beta , \, \D_\gamma ] = 0$.
We would like to mention that this algebra is already known \cite{K}.
\vspace{0.2cm}

Note that the algebras $A_{I}$ and $A_{II}$ can directly be
extended to the case $i,j = 0,1,2 \dots , N-1$ for $N >3$ (they correspond
to the algebras $A_I(N)$ and $A_{II}(N)$ discussed in Sec.4).

\vspace{0.2cm}
\noindent
{\bf Case B.} \underline{Among the coefficients $x_\alpha$, $x_\beta$ and
$x_\gamma$ one is equal to zero.}\vspace{3mm}

Let $x_\alpha , x_\gamma \neq 0\ , \; x_\beta = 0$.
In this case,
the Eqs.(\ref{PBW2}), (\ref{PBW4}), (\ref{PBW6}) become trivial and
Eq.(\ref{PBW5}) gives
\be
\lb{b1}
g_{\beta\alpha} = g_{\gamma\beta}\ .
\ee
There are two ways to satisfy the remaining Eqs.(\ref{PBW1}) and
(\ref{PBW3}).

\vspace{0.2cm}
\noindent
{\bf 1).}~
Eqs.(\ref{PBW1}) and (\ref{PBW3}) are satisfied if one chooses
$\Lambda_{\alpha\gamma} = \Lambda_{\alpha\beta} = \Lambda_{\beta\gamma}=:\Lambda$
which, in view of
(\ref{b1}), leads to
\be
\lb{b2}
g_{\alpha\beta} = g_{\beta\gamma}\ , \qquad
g_{\gamma\alpha} = g_{\alpha\gamma} + g_{\beta\alpha} - g_{\alpha\beta}\ .
\ee
The corresponding  algebra is
\be
\lb{b3}
{\bf B^{(1)}}\;\;
\mbox{\framebox[112mm]{$
\begin{array}{lcl}
g_{\beta}\ \D_{\alpha} \D_{\beta} -
(g_{\beta}-\Lambda)\ \D_{\beta} \D_{\alpha} &=&
 -  x_{\alpha}\D_\beta\ , \\[1mm]
g\  \D_{\alpha} \D_{\gamma}-
(g-\Lambda)\ \D_{\gamma} \D_{\alpha} &=&
x_\gamma \D_\alpha -  x_\alpha \D_\gamma\ , \\[1mm]
g_{\beta}\ \D_{\beta} \D_{\gamma} -
(g_{\beta}-\Lambda)\ \D_{\gamma} \D_{\beta} &=&
x_\gamma  \D_\beta\ ,\quad\qquad \forall\; g,\ g_\beta\neq 0\ .
\end{array}
$}}
\ee
Here for sake of future convenience we have parameterized
the bulk rates via $g_\beta$, $g$ and $\Lambda$. The algebra
(\ref{b3}) is also known \cite{ADR,AHR}.

\vspace{0.2cm}
\noindent
{\bf 2).}~ If $g_{\gamma\beta} = g_{\beta\alpha} = 0$, then
Eqs.(\ref{PBW1}), (\ref{PBW3}) are trivially satisfied and the
algebra reads
\be
\lb{b4}
{\bf \phantom{a} B^{(2)}} \;\;
\mbox{\framebox[112mm]{$
\begin{array}{l}
g_{\alpha \beta}\ \D_{\alpha} \D_{\beta} \; =\;
 -  x_\alpha \D_\beta\ , \\[1mm]
g_{\alpha \gamma} \,  \D_{\alpha} \D_{\gamma} -
g_{\gamma \alpha} \, \D_{\gamma} \D_{\alpha} \; =\;
x_\gamma \D_\alpha -  x_\alpha \D_\gamma\ , \\[1mm]
g_{\beta \gamma}\ \D_{\beta} \D_{\gamma}  \; =\;
x_\gamma  \D_\beta\ ,\qquad\qquad\qquad\quad \forall\;\;
g_{\alpha\beta},\ g_{\alpha\gamma},\ g_{\beta\gamma}\neq 0\ .
\end{array}
$}}
\ee
This algebra can be found already in Refs.\cite{ADR,AHR}.
Notice that if one takes $g_\beta =\Lambda$
in Eq.(\ref{b3}) one obtains a special case
of the algebra $B^{(2)}$ (Eq.(\ref{b4})).

\vspace{.2cm}

We will not repeat the same considerations
for the cases $x_\alpha, x_\beta\neq 0, x_\gamma=0~$
and $~x_\beta, x_\gamma\neq 0, x_\alpha=0$. The resulting
algebras are:

\vspace{0.2cm}
\noindent
\be
\lb{b5}
\phantom{a}\hspace{-4mm}
\left\{
\begin{array}{lcl}
g\ \D_{\alpha} \D_{\beta} -
(g-\Lambda)\ \D_{\beta} \D_{\alpha} &=&
x_\beta \D_\alpha -  x_\alpha \D_\beta\ , \\[1mm]
g_\gamma\  \D_{\alpha} \D_{\gamma}-
(g_\gamma-\Lambda)\ \D_{\gamma} \D_{\alpha} &=&
 -  x_\alpha \D_\gamma\ , \\[1mm]
(g_{\gamma}-\Lambda)\ \D_{\beta} \D_{\gamma} -
g_{\gamma}\ \D_{\gamma} \D_{\beta} &=&
-x_\beta  \D_\gamma\ ,\qquad\quad\;\; \forall\; g\neq 0\ ,\; g_\gamma\notin\{0,\Lambda\}\ ;
\end{array}
\right.
\ee
\noindent
\be
\lb{b6}
\phantom{a}\hspace{-4mm}
\left\{
\begin{array}{lcl}
(g_{\alpha}-\Lambda)\ \D_{\alpha} \D_{\beta} -
g_{\alpha}\ \D_{\beta} \D_{\alpha} &=&
x_\beta \D_\alpha\ , \\[1mm]
g_\alpha\  \D_{\alpha} \D_{\gamma} -
(g_\alpha-\Lambda)\ \D_{\gamma} \D_{\alpha} &=&
x_\gamma \D_\alpha\ , \\[1mm]
g\ \D_{\beta} \D_{\gamma} -
(g-\Lambda)\ \D_{\gamma} \D_{\beta} &=&
  x_\gamma\D_\beta - x_\beta\D_\gamma\ ,\;\; \forall\; g\neq 0\ ,\; g_\alpha\notin\{0,\Lambda\}\ ;
\end{array}
\right.
\ee
\vspace{2mm}

\noindent
\be
\lb{b7}
\phantom{a}\hspace{-37mm}
{\bf B^{(3)}} \;
\mbox{\framebox[105mm]{$
\begin{array}{l}
g\ \D_{\alpha} \D_{\beta}  -
(g-\Lambda)\ \D_{\beta} \D_{\alpha} \; =\;
x_\beta \D_\alpha -  x_\alpha \D_\beta\ , \\[1mm]
g_\gamma\  \D_{\alpha} \D_{\gamma} \; =\;
 -  x_\alpha \D_\gamma\ , \\[1mm]
(g_{\gamma}-\Lambda)\ \D_{\beta} \D_{\gamma}  \; =\;
-  x_\beta \D_\gamma\ ,\qquad \forall\; g\neq 0\ ,\; g_\gamma\notin\{0,\Lambda\}\ ;
\end{array}
$}}
\ee
\vspace{2mm}

\noindent
\be
\lb{b8}
\phantom{a}\hspace{-4.5mm}
{\bf B^{(4)}}\;
\mbox{\framebox[127mm]{$
\begin{array}{l}
(g_{\alpha}-\Lambda)\ \D_{\alpha} \D_{\beta} \; =\;
x_\beta \D_\alpha\ , \\[1mm]
g_{\alpha}\  \D_{\alpha} \D_{\gamma} \; =\;
x_\gamma \D_\alpha\ ,  \\[1mm]
g\ \D_{\beta} \D_{\gamma} -
(g-\Lambda)\ \D_{\gamma} \D_{\beta} \; =\;
x_\gamma \D_\beta - x_\beta \D_\gamma\ ,\;\forall\; g\neq 0\ ,\; g_\alpha\notin\{0,\Lambda\}\ .
\end{array}
$}}
\ee

\vspace{2mm}
\noindent
The algebras (\ref{b5}) and (\ref{b6}) are
just different presentations of the algebra $B^{(1)}$ (one has
to make the substitution $\beta\leftrightarrow\gamma$
in Eq.(\ref{b5}) and $\alpha\leftrightarrow\beta$
in Eq.(\ref{b6})).

The relation between the algebras $B^{(3)}$ and $B^{(4)}$ is less
trivial (therefore we keep them as different cases in classification).
One can obtain the relations for the $B^{(3)}$ algebra by inverting the order of
all products in the $B^{(4)}$ algebra (Eqs.(\ref{b8})), i.e.
by reading the relations (\ref{b8}) from the right to the left
and changing the signs of all $x$'s.
That means, the algebras $B^{(3)}$ and $B^{(4)}$
describe mirror (left-right) symmetric physical processes.

The algebras $B^{(3)}$ and $B^{(4)}$ are
completely different from
$B^{(2)}$. Even the number of independent rates in cases
$B^{(3)}$ and $B^{(4)}$ is not the same as in $B^{(2)}$.
\vspace{0.2cm}

\noindent
{\bf Case C.} \underline{Two of the coefficients $x_\alpha$, $x_\beta$ and
$x_\gamma$ are equal to zero.}\vspace{3mm}

In this case only one of the Eqs.(\ref{PBW1})--(\ref{PBW6})
remains nontrivial and the analysis becomes straightforward.
Below we present the relations for the type $C$ algebras
with $x_\alpha\neq 0$, $x_\beta=x_\gamma=0$. The expressions for the cases
$x_\beta\neq 0$, $x_\alpha=x_\gamma=0$ and
$x_\gamma\neq 0$, $x_\alpha =x_\beta=0$ can be obtained by
the substitutions $\alpha\leftrightarrow \beta$ and, respectively,
$\alpha\rightarrow\gamma\rightarrow\beta\rightarrow\alpha$ in
(\ref{c1}) and (\ref{c2}).

\vspace{0.2cm}
\noindent
\be
\lb{c1}
{\bf C^{(1)}} \;
\mbox{\framebox[102mm]{$
\begin{array}{l}
g_\beta\ \D_{\alpha} \D_{\beta} -
(g_\beta-\Lambda)\ \D_{\beta} \D_{\alpha} \; =\; - x_\alpha \D_\beta\ , \\[1mm]
g_\gamma\ \D_{\alpha} \D_{\gamma} -
(g_\gamma-\Lambda)\ \D_{\gamma} \D_{\alpha} \; =\;
 - x_\alpha \D_\gamma\ , \\[1mm]
g_{\beta \gamma} \, \D_{\beta} \D_{\gamma} -
g_{\gamma \beta} \, \D_{\gamma} \D_{\beta} \; =\; 0\ ,
\qquad\quad\forall\; g_\beta, g_\gamma,
g_{\beta\gamma}\neq 0\ ;
\end{array}
$}}
\ee
\noindent
\be
\lb{c2}
{\bf C^{(2)}} \;
\mbox{\framebox[102mm]{$
\begin{array}{l}
g_{\alpha \beta} \, \D_{\alpha} \D_{\beta} -
g_{\beta \alpha} \, \D_{\beta} \D_{\alpha} \; =\; -  x_\alpha \D_\beta\ , \\[1mm]
g_{\alpha \gamma} \, \D_{\alpha} \D_{\gamma} -
g_{\gamma \alpha} \, \D_{\gamma} \D_{\alpha} \; =\; -  x_\alpha \D_\gamma\ , \\[1mm]
g_{\beta\gamma}\ \D_{\beta} \D_{\gamma} \; =\; 0\ ,\qquad\qquad
\qquad\quad\;
\forall\; g_{\alpha\beta}, g_{\alpha\gamma}, g_{\beta\gamma}\neq 0\ .
\end{array}
$}}
\ee

\vspace{2mm}
\noindent
We observe that for $g_{\gamma\beta}=0$,
the algebra (\ref{c1}) is a special case
of the algebra (\ref{c2}). For the sake of convenience
(see Sec.4), we will
not stress this observation any further.

Notice that in case $\Lambda\neq 0$
shifting the generator $\D_\alpha$:
$\D_\alpha \rightarrow
\D_\alpha - {x_\alpha}/{\Lambda}$,
in Eq.(\ref{c1}) one obtains
\be
\lb{c1a} \left\{
\begin{array}{l}
g_\beta\ \D_{\alpha} \D_{\beta} -
(g_\beta-\Lambda)\ \D_{\beta} \D_{\alpha} \; = \; 0 \ , \\[1mm]
g_\gamma\ \D_{\alpha} \D_{\gamma} -
(g_\gamma-\Lambda)\ \D_{\gamma} \D_{\alpha} \; =\; 0 \ , \\[1mm]
g_{\beta \gamma} \, \D_{\beta} \D_{\gamma} -
g_{\gamma \beta} \, \D_{\gamma} \D_{\beta} \; =\; 0\ ,
\end{array}
\right.
\ee
which brings the $C^{(1)}$ algebra to
the subcase of a family of quantum hyperplanes (see case $D$
below)\footnote{ Conversely, using the linear shifts of
generators $\D'_i = \D_i + u_i$,
$\forall\; i\in\{\alpha , \beta, \gamma\}$ in the $D$-type
algebra (\ref{d}) and demanding the
resulting relations to  agree with the diffusion algebra Ansatz
(\ref{algebra}) one  recovers the $C^{(1)}$-type algebras only. }.
If one thinks of applications to stochastic processes, the shift we
made is not an innocent one
since for $g_{\beta\gamma} = g_{\gamma\beta}$, the algebra (\ref{c1})
has representations with traces (for example, the one-dimensional representation given by the shift) whereas the algebra (\ref{c1a}) has none.

A useful algebra (not of PBW type) is obtained
if one takes not only $g_{\gamma\beta}=0$ but also $g_{\beta\gamma}=0$ in
the algebra $C^{(2)}$ given by Eq.(\ref{c2}). In this way one
obtains the algebra used in Ref.\cite{E}. \vspace{2mm}

\noindent
{\bf Case D.} \underline{All the coefficients $x_\alpha$, $x_\beta$ and
$x_\gamma$ are equal to zero.}\vspace{3mm}

In this case we obtain the algebra of Manin's quantum hyperplane \cite{M}
corresponding to multiparametric Drinfeld-Jimbo R-matrix \cite{DJ}
\be
\lb{d}
{\bf D} \;\;
\mbox{\framebox[125mm]{$
g_{ab} \, \D_a \D_b -
g_{ba} \D_a \D_b = 0\ ,
\quad \forall \; a, b\in
\{\alpha,\beta,\gamma\}\ :\;\;
a<b\ , \;\;
 g_{ab}\neq 0\ .$}}
\ee

\vspace{2mm} A different point of view in understanding some of
the algebras presented here is discussed in Appendix A. \vspace{3mm}

\section{Diffusion algebras with $N > 3$ generators.}
\setcounter{equation}0

While classifying  the PBW-type algebras (\ref{algebra}) with
more then 3 generators one meets  the combinatorial problem of
consistently combining several minimal subalgebras generated by
triples $\{\D_\alpha, \D_\beta, \D_\gamma\}$ (each of these
subalgebras belonging to one of the types $A$--$D$ listed in Sec.3)
to a larger algebra. In this section we shall first construct
several basic series of diffusion algebras being extensions of
the $A_I$, $A_{II}$, $B^{(1)-(4)}$, $C^{(1)}$ and $D$ -type algebras from
the previous section. The $C^{(2)}$-type
triples will appear later on in our considerations.
There is a deep reason behind our choice of starting first with the
algebra $C^{(1)}$ and taking into account the algebra $C^{(2)}$
later. In this way
one can easier state the theorem presented at the
end of this section and which is the central part
of our work.
A proof of this theorem will be given elsewhere.

In Sec.3 we have shown that the classification of diffusion
algebras with $N=3$ generators depends essentially on the number of
nonvanishing parameters $x_\alpha$ in the Ansatz (\ref{algebra}).
The same is true for general $N$. Therefore we shall split the
set $\{\alpha\}$ labeling different species of particles ($=$
different generators $\D_\alpha$) into two subsets
$\{\alpha\}$=$\{i\}\cup\{a\}$. From now on we
assign letters $i$, $j$, $k$, etc. to the indices of
the first subset and assume that $x_i, x_j, x_k, \dots \neq 0$.
The indices of the second subset are denoted by letters $a$, $b$,
$c$, etc. and it is implied that $x_a=x_b=x_c=\dots =0$. Let
$N_1$ and $N_0$ denote the number of elements  of
the first and second subsets. Clearly, $N_0 + N_1 = N$ ---
the total number of indices of both kinds.

We should stress however that $N_1$ --- the number of nonzero
$x$'s --- is the most noticeable but not the only relevant information
 for the
classification. A supplementary information is given by a number of
nonzero bulk rates $g_{\alpha \beta}$
in the defining relations of the algebras
(cf. cases $A_I$ and $A_{II}$, $B^{(1)}$ and
$B^{(2)}$, or $C^{(1)}$ and $C^{(2)}$ from Sec.3) and the mutual
arrangement of the indices $\{i\}$ and $\{a\}$ in the
alphabetic order (see Eq.(\ref{order}) and the definition of the algebras $B^{(2)}$, $B^{(3)}$ and
$B^{(4)}$ from Sec.3).

\vspace{3mm}
\noindent
{\bf Algebras of type A.} We shall start by considering the algebras
with the number $N_1$ of nonzero $x$'s
not less than $3$ --- we call them algebras of type $A$.
Obviously, any such algebra contains a minimal subalgebra of
type $A_I$, or $A_{II}$. So, these algebras are naturally
obtained by a sequence of consistent (in a sense of diamond
conditions) extensions starting with the $N=N_1=3$ algebras
(\ref{a5}) or (\ref{a6}) and adding one new generator
$\D_\alpha$ at each step of the iteration.

It is suitable to begin the extension procedure with the generators
whose indices lie in the subset $\{i\}$.
At the first step one adds a fourth
generator, say $\D_l$, to the triple $\{\D_i,\D_j,\D_k\}$
(recall, once again, that $x_i, x_j, x_k, x_l \neq 0$).
The resulting algebra contains four minimal subalgebras
of the types either $A_I$, or $A_{II}$.
An easy check shows that it is possible to combine only triples of the
same type. Continuing the extension procedure one finally obtains
algebras with $N_1$ generators for which all the minimal subalgebras are
of the same type, either $A_I$ or $A_{II}$. The defining relations for
these two types of diffusion algebras --- $A_I(N_1)$ and $A_{II}(N_1)$ ---
are given by the Eqs.(\ref{a5}) and, respectively (\ref{a6}), with
$i<j$ spanning the whole set $\{i\}$.

We continue the extension procedure adding to the algebras $A_{I,II}(N_1)$,
$N_0$ new generators with their indices lying in the subset $\{a\}$.
First, we add one generator, say $\D_a$, and take care that all the
newly appeared triples $\{\D_a,\D_i,\D_j\}$ belong to one of the algebras
of type $B$ (see
Eqs.(\ref{b3}), (\ref{b4}), (\ref{b7}) and (\ref{b8})).
Next, adding a second generator, say $\D_b$, we
again require that all the triples $\{\D_b,\D_i,\D_j\}$ are of type
$B$ and, moreover, demand that the triples $\{\D_a,\D_b,\D_i\}$
are the generators of a $C_I$-type algebra (\ref{c1}). Adding new
generators
we have also to impose $D$-type algebraic relations
for the triples $\{\D_a, \D_b, \D_c\}$.
As a result,
we obtain two different extensions for the algebra $A_I(N_1)$

\vspace{0.2cm}
\noindent
\be
\lb{a1a}
\phantom{a}\hspace{-6mm}
{\bf A}_{\bf I}^{(1)}(N_1,N_0) \;\;
\mbox{\framebox[116mm]{$
\begin{array}{l}
g \, [ \D_i , \, \D_j ]  = x_j \, \D_i -  x_i \, \D_j\ ,\;\;\, \forall \;
i,j \in \{i\}: \; i\neq j; \;\; g\neq 0\ , \\[1mm]
g_a \, [ \D_a , \, \D_i ]  = x_i \, \D_a\ ,
\qquad\quad\;\;  \forall i\in \{i\},\;
\forall a\in\{a\}; \;\; g_a\neq 0\ , \\[1mm]
g_{ab} \, \D_{a} \D_{b} - g_{ba} \D_{b} \D_{a} = 0 \ ,
\quad\; \forall \; a,b \in \{a\}: \; a<b; \;\;  g_{ab}\neq 0\ .
\end{array}
$}}
\ee

\vspace{0.2cm}
\noindent
\be
\lb{a1b}
\phantom{a}\hspace{-5mm}
{\bf A}_{\bf I}^{(2)}(N_1,N_0) \;
\mbox{\framebox[116mm]{$
\begin{array}{l}
g \, [ \D_i , \, \D_j ]  = x_j \, \D_i -  x_i \, \D_j\ ,\;\; \forall \;
i,j \in \{i\}: \; i\neq j; \;\; g\neq 0\ , \\[2mm]
g_+ \, \D_i \, \D_b  = - x_i \,
\D_b\ , \;\;\quad\;\; \forall ( i \in \{i\}, \; b  \in\{a\}): \; i < b ;
\;\; g_+\neq 0 , \\[2mm]
g_- \, \D_a \, \D_i  =  x_i \, \D_a\ , \qquad\;\;
\forall ( i \in \{i\}, \; a  \in\{a\}): \;
i > a ; \;\; g_-\neq 0\, , \\[2mm]
\mbox{where}\;\; g_-=-g_+\ ,
\;\; \mbox{if there exist} \;\; D_i, D_a, D_b\, :\quad a<i<b\ , \\[2mm]
g_{ab} \, \D_{a} \D_{b} - g_{ba}
\D_{b} \D_{a} = 0 \ , \;\; \forall\; a,b \in \{a\}: \ a<b; \;\; g_{ab} \neq 0\ .
\end{array}
$}}
\ee

\vspace{1mm}
\noindent
Here and in what follows we indicate two integers $N_0$ and $N_1$ ---
the numbers of indices in the sets $\{i\}$ and $\{a\}$, respectively ---
in braces to specify the type of algebra.
As the reader can notice looking closely at Eq.(\ref{a1b}), specifying
$N_0$ and $N_1$ one obtains several algebras all denoted by
$A_I^{(2)}(N_1,N_0)$.
We didn't introduce a different notation for each algebra in order to
simplify the notations. We adopted the same attitude also for other
algebras described below (see Eqs.(\ref{a6a}) and (\ref{b2a})).
\medskip

The algebra $A_{II}(N_1)$
possesses a unique extension

\vspace{5mm}
\noindent
\be
\lb{a6a}
\phantom{a}\hspace{-5mm}
{\bf A}_{\bf II}(N_1,N_0) \;\;
\mbox{\framebox[116mm]{$
\begin{array}{l}
 g_{ij} \, \D_i \, \D_j  =
x_j \, \D_i -  x_i \, \D_j\
 ,\quad\; \forall \;
i,j \in \{i\}: \; i< j \ , \\[2mm]
g_{i+} \, \D_i \, \D_b  = - x_i \,
\D_b\ , \;\;\quad\qquad
\forall ( i \in \{i\}, \; b  \in\{a\}): \; i < b , \ , \\[2mm]
g_{i-} \, \D_a \, \D_i  =  x_i \, \D_a\ , \qquad\qquad\,
\forall ( i \in \{i\}, \; a  \in\{a\}): \;
i > a \ . \\[2mm]
\mbox{Here}\quad g_{ij}:=g_i-g_j\ ,\quad g_{i+}:=g_+ +g_i\ ,
\quad g_{i-}:=g_--g_i\ , \\[1mm]
\mbox{where for all $i$:}\;\;\; g_i\neq -g_+\ , \;\ g_i\neq g_-\ ;
\quad g_i\neq g_j\quad \mbox{if}\quad i\neq j \ ;
 \\[1mm]
\mbox{and}\;\; g_-=-g_+\ ,
\quad \mbox{if there exists}\;\; D_i, D_a, D_b\, :\quad a<i<b\ . \\[2mm]
g_{ab} \, \D_{a} \D_{b} - g_{ba}
\D_{b} \D_{a} = 0 \ , \;\; \forall\; a,b \in \{a\}: \ a<b;
\;\; g_{ab}\neq 0\ .
\end{array}
$}}
\ee

\vspace{2mm}
The parameters $g_+$ and $g_-$ in algebras $A_I^{(2)}(N_1,N_0)$ and $A_{II}(N_1,N_0)$
remain independent provided that mutual order of indices
of the subsets $\{i\}$ and $\{a\}$
(and hence the order of the generators $D_i$ and $D_a$, see (\ref{order}))
 is like follows
$$
i_1<i_2<\dots <i_k\ <\ a_1<a_2<\dots<a_{N_0}\ <\ i_{k+1}<i_{k+2}<\dots <i_{N_1}\ .
$$
Only in this case all the $B$-type minimal subalgebras in the algebras (\ref{a1b})
and (\ref{a6a}) belong to  the type $B^{(2)}$. In the presence of $B^{(3)}$, or
$B^{(4)}$ type triples the parameters $g_+$ and $g_-$ are constrained by
condition
$g_+ + g_- =0$
(keep in mind that for stochastic processes all rates have to be non-negative).

\vspace{4mm}
\noindent
{\bf Algebras of type B.} We now consider the case where the set $\{i\}$
contains exactly two indices ($N_1=2$), say, $i$ and $j$, $i<j$.
To obtain such algebras --- we call them the algebras of type $B$ ---
one should consider consistent extensions of the $B$-type triples
(\ref{b3}), (\ref{b4}), (\ref{b7}) and (\ref{b8})  by
$(N_0-1)$ generators with their labels in the set $\{a\}$.

Starting with the $B^{(1)}$-type triple $\{\D_i,\D_{a_1},\D_j\}$,
the only possibility  is to add new generators $\D_{a_2}$, $\D_{a_3}$, etc.
such that all the minimal subalgebras $\{\D_i,\D_{a_2},\D_j\}$,
$\{\D_i,\D_{a_3},\D_j\}$, etc.
satisfy again $B^{(1)}$-type relations (\ref{b3}).
In this situation, we can arrange the alphabetic order of the generators
as follows:
$\D_j>\D_{a_1}>\D_{a_2}>\dots>\D_{a_{N_0}}>\D_i$
(cf. with the comment below Eq.(\ref{b8})). So it is natural to put in
this case $i=0$, $j=N-1$. The extended algebra ${\bf B^{(1)}}(2,N_0)$
reads\nopagebreak

\vspace{0.3cm}
\noindent
\be
\lb{b1a}
\mbox{\framebox[136mm]{$
\begin{array}{l}
 g\  \D_0\  \D_{N-1} - (g-\Lambda)\ \D_{N-1}\ \D_0\ =
x_{N-1}\  \D_0 -  x_0\ \D_{N-1}\ , \qquad\qquad\; g\neq 0\ ,
\\[2mm]
g_a\  \D_0\  \D_a -(g_a-\Lambda)\ \D_a\ \D_0 = - x_0\
\D_a\ ,  \\[2mm]
g_a\  \D_a\  \D_{N-1} -(g_a-\Lambda)\ \D_{N-1}\ \D_a =  x_{N-1}\
\D_a\ , \;\;  \forall\; 1<a<N\!-\! 1 ,\; g_a\neq 0\ ,
\\[2mm]
g_{ab}\  \D_{a}\ \D_{b} - g_{ba}\
\D_{b}\ \D_{a} = 0 \ , \qquad\qquad\qquad\;
 \forall\; a,b \in \{a\}: \ a<b; \;\;\, g_{ab}\neq 0\ .
\end{array}
$}}
\ee

\vspace{2mm}
The situation becomes
different for the other $B$-type triples (\ref{b4}), (\ref{b7}) and (\ref{b8}).
Extending these algebras
one can get algebras containing all the $B^{(2)}$, $B^{(3)}$ and $B^{(4)}$-type
minimal subalgebras. Therefore we introduce a unified notation
$B^{(2)}(2,\ n_{<},n,n_{>})$ for extensions of the triples $B^{(2)}-B^{(4)}$.
Here $n_{<} +n + n_{>} =N_0$
and a mutual order of the indices $i$, $j$ and the indices from the set $\{a\}$
is as follows
$$
a_1<\dots <a_{n_{<}}<i<a_{(n_{<}+1)}<\dots
<a_{(n_{<}+n)}<j<a_{(n_{<}+n+1)}<\dots <a_{N_0}\ .
$$
The algebra ${\bf B^{(2)}}(2,\ n_{<},n,n_{>})$
reads

\vspace{2mm}
\noindent
\be
\lb{b2a}
\phantom{a}\hspace{-2mm}
\mbox{\framebox[136mm]{$
\begin{array}{l}
 g \, \D_i \, \D_j - (g-\Lambda)\ \D_j\ \D_i\ =
x_j \, \D_i -  x_i \, \D_j\ , \qquad\qquad\qquad\quad g\neq 0\ ,
\\[2mm]
g_+ \, \D_i \, \D_a  = - x_i\ \D_a\ ,\;\;\qquad\;\;
(g_+ - \Lambda)\ \D_j\
\D_a = -x_j\ \D_a\ , \;\; \forall \;\; a>j\ ,
\\[2mm]
(g_- - \Lambda) \, \D_a
\, \D_i  =  x_i \, \D_a\ , \quad\;\; g_-\ \D_a\ \D_j = x_j\ \D_a\ ,
\qquad\qquad  \forall\;\; a< i\ ,
\\[2mm]
g_+\ \D_i\ \D_a = -x_i\ \D_a\
,\qquad\;\;\;\, g_-\ \D_a\ \D_j = x_j\ \D_a\ , \qquad\quad\;\;\;
\forall \;\; i<a<j\ ,
\\[2mm]
\mbox{where}\quad g_+\neq \left\{
\begin{array}{lcl}
0 &\mbox{if}& n_<<N_0 \\
\Lambda &\mbox{if}& n_>>0
\end{array}
\right.  , \quad
g_-\neq
\left\{
\begin{array}{lcl}
0 &\mbox{if}& n_><N_0 \\
\Lambda &\mbox{if}& n_<>0
\end{array}
\right.  ,
\\[4mm]
\mbox{and} \; g_+ + g_- = \Lambda \; \mbox{if among the numbers}\;
n_<, n, n_>\; \mbox{there are two nonzeros} ,
\\[3mm]
g_{ab} \, \D_{a} \D_{b} - g_{ba}
\D_{b} \D_{a} = 0 \ , \qquad\qquad\;\;\;\; \forall\;
a,b \in \{a\} : \; a<b; \;\; g_{ab}\neq 0\ .
\end{array}
$}}
\ee
\vspace{1mm}

\noindent
The algebra $B^{(2)}(2,\ n_{<},n,n_{>})$ contains $n_{<}$ $B^{(4)}$-type
minimal subalgebras ($\{\D_a,\D_i,$ $\D_j\}$ for $a<i$),
$n$~ $B^{(2)}$-type minimal subalgebras ($\{\D_i,\D_a,\D_j\}$ for $i<a<j$)
and $n_>$ $B^{(3)}$-type triples ($\{\D_i,\D_j,\D_a\}$ for $a>j$).
\vspace{5mm}

\noindent {\bf Algebras of type C.} Next, we consider algebras with
only one
index $i$ in the subset $\{i\}$, let us call them the algebras of
the type $C$. These algebras arise from extension of the $C_I$
triple $\{\D_i, \D_a, \D_b\}$ by $(N_0-2)$ generators $\D_c,
\dots$ labeled by indices from the subset $\{a\}$. Checking the
consistency of such an extension is  straightforward and,
therefore,  we shall just present directly the resulting algebra

\vspace{0.3cm}
\noindent
\be
\lb{c1b}
{\bf C}(1,N_0) \;
\mbox{\framebox[115mm]{$
\begin{array}{l}
 g_a \, \D_i \, \D_a - (g_a-\Lambda)\ \D_a\ \D_i\ =
 -  x_i \, \D_a\  , \\[2mm]
\mbox{where}\;\;\quad g_a\neq 0\;\; \mbox{if}\;\; i<a\ , \quad\;\;\mbox{and}\;\;\quad
g_a\neq\Lambda\;\; \mbox{if}\;\; a<i\ ,
\\[2mm]
g_{ab} \, \D_{a} \D_{b} - g_{ba}
\D_{b} \D_{a} = 0 \ , \;\;\;\; \forall\;
a,b \in \{a\} : \; a<b; \;\; g_{ab}\neq 0\ .
\end{array}
$}}
\ee
\vspace{1mm}

\noindent
As in case of the $C^{(1)}$-type triple (\ref{c1}),  for $\Lambda\neq 0$
one can reduce the algebra (\ref{c1b}) to a subcase of the family of
quantum hyperplanes (see case $D$ below) shifting the
generator $\D_i$: $\D_i \rightarrow \D_i - x_i / \Lambda$.
Nevertheless, as will be shown in the theorem given below, it is useful
to keep the definition given by Eq.(\ref{c1b}) for the algebra
$C(1,N_0)$ since in
this way one can use it as a building block for the construction of new
algebras. In the new algebras the shift will not be possible anymore.

Note that, unlike all the previous cases, one can consider the algebra
$C(1,1)$ produced by a pair of generators. We shall use this possibility in
the theorem stated below. For instance, the $C^{(2)}$ algebra
(\ref{c2}) can be constructed as a combination of two $C(1,1)$ algebras
by a blending procedure described in the theorem given below.
Further examples of an application of $C(1,1)$ algebras are given in
Appendix B (see the cases 17 and 18).
\vspace{5mm}

\noindent
{\bf Case D.}
The $D$ algebras --- the algebras with no indices in the subset $\{i\}$ ---
are represented by a family of $N=N_0$ dimensional quantum hyperplanes

\vspace{0.3cm}
\noindent
\be
\lb{d1b}
{\bf D}(0,N_0) \;
\mbox{\framebox[115mm]{$
\begin{array}{l}
g_{ab} \, \D_{a} \D_{b} - g_{ba}
\D_{b} \D_{a} = 0 \ , \;\;\;\;\forall\;
a,b \in \{a\} : \; a<b; \;\;\, g_{ab}\neq 0\ .
\end{array}
$}}
\ee

Now we are ready to complete a classification scheme.
To do so one needs
to take into
consideration the possibility of using $C^{(2)}$-type
triples in the algebra extension process.
As a result one derives a procedure
to obtain all the
the diffusion algebras which is described in
the following theorem.

\medskip
\noindent {\bf Theorem.~}
{\em
A diffusion algebra has
$N_1$ generators $\D_i$, where one assumes
$x_i\neq 0$,
($i=1,2,\dots ,N_1$) and $N_0$ generators $\D_a$, with $x_a=0$,
($a=1,2,\dots ,N_0$). If we don't
distinguish between the two kinds of generators, we denote them by
$\D_\alpha$, ($\alpha=1,2,\dots ,N=N_0+N_1$).

If $N_1=0$, the algebras are $D(0,N_0)$
(see Eq.(\ref{d1b})).

If $N_1\neq 0$, all diffusion
algebras can be obtained by a blending procedure using the algebras
(\ref{a1a})--(\ref{c1b}).
The blending procedure can be described as follows.

Consider two of the
algebras (\ref{a1a})--(\ref{c1b}), denoted by $X(N_1,N_0^{(x)})$
and $Y(N_1,N_0^{(y)})$, both
having the same number of generators $\D_i$ which satisfy the same
relations among themselves in the two algebras, and generators $\D_{a_x}$
(respectively $\D_{a_y}$).
Through blending, one can obtain a new diffusion
algebra $Z(N_1,N_0^{(x)}+N_0^{(y)}=N_0)$ with generators $\D_i$ and
$\D_a$, ($a=1,2,\dots ,N_0$).
Since the
$X(N_1,N_0^{(x)})$
and $Y(N_1,N_0^{(y)})$
algebras are both of PBW type, the $N_0^{(x)}$ indices
(respectively the $N_0^{(y)}$ indices)
are in a given alphabetic order. We
blend now the $N_0^{(x)}$ and $N_0^{(y)}$ indices together in an arbitrary
alphabetic
order but respecting the order for the $N_0^{(x)}$ indices
(respectively the $N_0^{(y)}$
indices) which are fixed in the algebra $X(N_1,N_0^{(x)})$
(respectively $Y(N_1,N_0^{(y)})$). The
alphabetic order of the $N_0^{(x)}$ indices in respect to
the $N_1$ indices (respectively the $N_0^{(y)}$
indices in respect to the $N_1$ indices)
given again by the two algebras $X(N_1,N_0^{(x)})$
and $Y(N_1,N_0^{(y)})$, has also to be respected.
For each alphabetic order of the $N_0$ indices one obtains a new
algebra in the following way. The relations among the generators
$\D_i$ and
those among the generators $\D_i$ and $\D_{a_x}$ (respectively
among $\D_i$ and $\D_{a_y}$) coincide with the relations in the algebras
$X(N_1,N_0^{(x)})$
and $Y(N_1,N_0^{(y)})$.
The remaining relations among the generators $\D_{a_x}$ and
$\D_{a_y}$ are:
\be
\lb{lacking}
\begin{array}{ll}
\D_{a_{x}} \D_{a_{y}} = 0\ , \qquad &\forall\;
a_x, a_y: \;\; a_{x}<a_{y}\ ,
\\
\D_{a_{y}} \D_{a_{x}} = 0\ , \qquad &\forall\;
a_x, a_y: \;\; a_{x}>a_{y}\ .
\end{array}
\ee
A "blended" algebra can now be blended with one of the algebras
(\ref{a1a})--(\ref{c1b}) and one can obtain a new algebra.
}
\vspace{2mm}

It is important to stress that through the blending procedure
one can obtain the same algebra using different blendings.
Therefore what we have is a construction rather than a classification of
the diffusion algebras.

We would like to point out that it is easy to see that one can blend
together only algebras of the same type: $A_I^{(1)}$ or
$A_I^{(2)}$ with
algebras $A_I^{(1)}$ or
$A_I^{(2)}$ , $A_{II}$ with $A_{II}$, $B^{(1)}$ or $B^{(2)}$ with
$B^{(1)}$ or $B^{(2)}$.
and $C$ with $C$.
One can show that blending together two $A_I^{(1)}$
(respectively $B^{(1)}$)
algebras leads to $A_I^{(1)}$
(respectively $B^{(1)}$) algebras. Therefore one can use
an $A_I^{(1)}$ (respectively $B^{(1)}$)
algebra only once during blending procedure. At
the same time the algebras $A_I^{(2)}$
(respectively $B^{(2)}$) can be blended any
number of times.

Let us show an application of this theorem. Consider the
algebra
\begin{eqnarray}
\lb{xx}
&&g_{0a}\  \D_0\ \D_a\ -g_{a0}\ \D_a\ \D_0  =  - x_0\ \D_a\ , \qquad
a=1,2,\dots ,N-1\ ,
\\[2mm]
\lb{xxx}
&& g_{ab}\ \D_a\ \D_b\ = 0\ , \qquad\qquad\qquad\quad\; \forall\; a,b:\; 1 \leq a<b \leq N-1\ .
\end{eqnarray}
This algebra is obtained taking $N-1$ copies of the algebra $C(1,1)$ ($\{i\}=\{0\}$,
$\{a\}=\{1,2,..,N-1\}$) and blending them together.
Taking $N=3$ and $4$ one recovers
the algebras given by Eq.(\ref{c2}) respectively
Eqs.(\ref{B18}), (\ref{B18a}).
If in Eq.(\ref{xxx}) one
takes the rates $g_{ab}=0$ one recovers the algebra discussed in Ref.\cite{E}.

\section{Discussion.}
\setcounter{equation}0

We have defined diffusion algebras. Those are PBW algebras with $N$
generators satisfying the relations (\ref{algebra}).
These algebras are
useful to find
stationary states of the stochastic processes given by the rates
$g_{\alpha\beta}$. For the N species problem one finds several series of algebras
which might be useful in applications using the
Eq.(\ref{1.3}) for a ring
and the Eqs.(\ref{1.7}) and
(\ref{1.8}) for open systems. Much work is still left. For
example one has to find which boundary conditions, if any, are compatible
with each of the algebras.
We would like to stress that all the cases which were used up to now
for applications are special cases of our construction.

An open and
relevant question is: do non PBW type
"physically meaningful"
algebras satisfying Eq.(\ref{algebra}) exist.\footnote{
These algebras can be finite or infinite dimensional.}
Such algebras
could eventually represent "exceptional algebras" similar to those which
appear in the theory of simple Lie algebras or superalgebras.

An interesting and unsolved problem is the connection between the diffusion
algebras and the so called quantum Lie algebras (see e.g. \cite{BIO})
which look
similar and which are not fully investigated and classified.
In order to show that such a connection might be possible we will give
here two examples of quantum Lie algebras related to the so-called
Cremmer-Gervais R matrix \cite{CG}:
\be
\lb{5.1}
\phantom{a}\hspace{3mm}
\left\{
\begin{array}{lcl}
g_{\beta}\ \D_{\alpha} \D_{\beta} -
(g_{\beta}-\Lambda)\ \D_{\beta} \D_{\alpha} &=&
 -  x_\alpha\D_\beta\ , \\
g\  \D_{\alpha} \D_{\gamma}-
(g-\Lambda)\ \D_{\gamma} \D_{\alpha} + w\ (\D_\beta)^2 &=&
x_\gamma\D_\alpha -  x_\alpha\D_\gamma\ , \\
g_{\beta}\ \D_{\beta} \D_{\gamma} -
(g_{\beta}-\Lambda)\ \D_{\gamma} \D_{\beta} &=&
x_\gamma \D_\beta\ ,
\end{array}
\right.
\ee
and
\be
\lb{5.2}
\left\{
\begin{array}{l}
g_{\alpha\beta}\ \D_{\alpha} \D_{\beta} \; =\;
 -  x_\alpha\D_\beta\ , \\
g_{\alpha\gamma}\  \D_{\alpha} \D_{\gamma}-
g_{\gamma\alpha}\ \D_{\gamma} \D_{\alpha} + w\ (\D_\beta)^2\; =\;
x_\gamma\D_\alpha -  x_\alpha\D_\gamma\ , \\
g_{\beta\gamma}\ \D_{\beta} \D_{\gamma}  \; =\;
x_\gamma\D_\beta\ ,
\end{array}
\right.
\ee
They are clearly extensions of the $B^{(1)}$ and
$B^{(2)}$ algebras to which  the term $(\D_\beta)^2$ has been
added ($w$ is an
arbitrary parameter). The algebras (\ref{5.1}) and (\ref{5.2})
are of PBW type. Quantum Lie algebras can also be relevant in a
different context un-related to stochastic processes. As pointed
out in Ref.\cite{AR}, quadratic algebras are useful also to
describe the ground-states of one-dimensional quantum chains in
equilibrium statistical physics if it happens
that the ground-states have energy zero. This is a whole area which
is worth exploring.

 Before closing this paper, we would like to mention a natural extension
of our results. The starting point of our investigation were the
processes given in Eq.(\ref{1.1}) which are related to quantum
hyperplanes (take all the $x$'s equal to zero in Eq.(\ref{algebra})).
One
can consider more general stochastic processes
in which the bulk rates are related to quantum superplanes \cite{M}.
In these cases one
obtains equations which generalize Eq.(\ref{algebra}) (see Ref.\cite{ADR}).
This would lead us to
something to which one could coin the name of "reaction-diffusion
algebras". For the time being this is not more than a nice
thought.

\subsection*{\bf Acknowledgements}

The authors thank M.Evans, O.Ogievetsky, I.Peschel,  G.M.Schuetz,
A.Sudbery, V.Tarasov and R.Twarock
for useful discussions.
We would like to thank the Heisenberg-Landau Foundation
for financial support.
The work of V.R. was partly supported by the grant INTAS OPEN 97-01312.
The work of A.I. and P.P. was 
supported in part by RFBR grant \# 00-01-00299.
One of the authors (P.P.) is grateful to the Max-Planck-Institut fu\"r
Mathematik in Bonn for its warm hospitality during his work on this paper.

\section*{Appendix A.~
Comments on diffusion algebras with three generators.}

\def\theequation{A.\arabic{equation}}
\setcounter{equation}0

We consider the relations (\ref{algebra}) in the case when we have only three
generators:
\be
\lb{A.1}
\left\{
\begin{array}{c}
g_{01} \D_{0} \D_{1} -
g_{10} \D_{1} \D_{0} =
x_1 \, \D_0 - x_0 \, \D_1 \; , \\
g_{20} \D_{2} \D_{0} -
g_{02} \D_{0} \D_{2} =
x_0 \, \D_2 - x_2 \, \D_0 \; , \\
g_{12} \D_{1} \D_{2} -
g_{21} \D_{2} \D_{1} =
x_2 \, \D_1 - x_1 \, \D_2 \; ,
\end{array}
\right.
\ee
and look for the case when the rates satisfy the condition:
\be
\lb{A.2}
\Lambda_{012} = g_{01} - g_{10} +  g_{12} - g_{21} +  g_{20} - g_{02} = 0
\ee
It is trivial to verify \cite{AHR} that the $1 \times 1$ matrices (c-numbers)
\be
\lb{A.3}
D_i = \frac{x_i}{f_i} \; , \;\;\; (i = 0,1,2) \; ,
\ee
where
\be
\lb{A.4}
f_1 = f_0 + g_{01} - g_{10} \;\; , \;\;\; f_2 = f_0 + g_{02} - g_{20} \; ,
\ee
verify the relations (\ref{A.1})
($f_0$ is an arbitrary parameter). One can use
(\ref{A.3}) in order to compute,
using Eq.(\ref{1.3}), the probability distribution on
a ring.
The probability distribution one obtains for the stationary state is
trivial (one has no correlations). The physics of the stationary state can
however be interesting if one takes an open system. We then have to use
Eq.(\ref{1.8}) and the one-dimensional representation of (\ref{A.1}) is not of much
help. If the rates satisfy only the condition (\ref{A.2}) it is not clear if one
doesn't have only the one-dimensional representation. At this point one
can understand why we are interested in algebras of PBW type. For
algebras of PBW type, we can be sure that one gets other representations
(at least the regular one). The price to pay is that we will get more
constraints on the rates than those given by Eq.(\ref{A.2}).
 In order to get algebras of PBW type compatible with the relation (\ref{A.2}),
it is useful to write the diamond condition (\ref{rel1}) in a different way:
\be
\lb{A.5}
\begin{array}{c}
  \Lambda_{012} \,
 \left( x_0 \, g_{21} \, D_2 \, D_1  +
x_1 \, g_{20} \, D_2 \, D_0  +
x_2 \, g_{21}\,  D_2 \, D_1
+ x_1 \, x_2   \, D_0
 - x_0 \, x_1  \,   D_2 \right)\hspace{5mm}
\\[2mm]
\phantom{a}\hspace{3mm}
+\ x_0 \, g_{12}\, g_{21}
\, [ D_1 ,\,  D_2]\  +\
x_1 \, g_{02}\, g_{20}
\, [ D_2 ,\, D_0 ]\ +\
x_2 \, g_{01}\, g_{10}
\, [ D_0 ,\, D_1 ]  = 0 \; .
\end{array}
\ee
We now take into account Eq.(\ref{A.2}). There are several solutions of
the Eq.(\ref{A.5}):
\be
\lb{A.6}
\phantom{a}\hspace{-37mm}
a) \quad\;\; g_{ij} = g\ .
\ee
This gives (using Eq.(\ref{A.1})) the algebra $A_I$ (see Eq(3.5)).
\be
\lb{A.7}
\phantom{a}\hspace{-10.5mm}
b) \;\;\;
\left\{
\begin{array}{l}
g_{21} = g_{20} = g_{10} = 0 \; , \\
g_{02} = g_{01} + g_{12}  \; .
\end{array}
\right.
\ee
This gives the algebra  $A_{II}$ (see Eq.(\ref{a6})).
\be
\lb{A.8}
\phantom{a}\hspace{-21mm}
c) \;\;\;
\left\{
\begin{array}{l}
 g_{20} = g_{10} = 0 \; , \\
\left[ \D_1 , \, \D_2 \right] = 0  \; .
\end{array}
\right.
\ee
This gives actually again the algebra $A_{II}$, defined in Eq.(\ref{a6}) with the
substitution
\be
\lb{A.10}
g_{12} - g_{21} \; \longrightarrow \; g_{12}\ .
\ee
This different derivation of the algebra $A_{II}$ has a bonus: we have
learned that we can take the quotient given by
the second relation in Eq.(\ref{A.8}) of this algebra.
\be
\lb{A.11}
d) \;\;\;
\left\{
\begin{array}{l}
x_0 = 0 \; , \\
 g_{01} = g_{02} = 0 \; , \\
g_{12} = g \; , \;\; g_{21} =  g - \Lambda \; , \\
g_{10} = g_\gamma  \; , \;\; g_{20} = g_\gamma - \Lambda  \; ,
\end{array}
\right.
\ee
which is the algebra $B^{(3)}$ (see Eq.(\ref{b7})).
\be
\lb{A.12}
e) \;\;\;
\left\{
\begin{array}{l}
x_0 = 0 \; , \\
 g_{10} = g_{20} = 0 \; , \\
g_{12} = g \; , \;\; g_{21} = g - \Lambda \; , \\
g_{01} = g_\alpha - \Lambda \; , \;\; g_{02} = g_\alpha  \; ,
\end{array}
\right.
\ee
which is the algebra $B^{(4)}$ (defined in Eq.(\ref{b8})).

Since the algebras $A_I$, $A_{II}$, $B^{(3)}$ and
$B^{(4)}$ are all derived starting from
Eq.(\ref{A.1}) with the conditions
(\ref{A.2}), all these algebras have at least a
one-dimensional representation. When the classification of the relations
(\ref{A.1}) for which a trace operation exists was done \cite{AHR}, there was
no need to
consider them separately since for the trace operation it is enough to
have the expressions (\ref{A.3}) and (\ref{A.4}).

\section*{Appendix B.~
Diffusion algebras with positive rates for $N=4$.}

\def\theequation{B.\arabic{equation}}
\setcounter{equation}0

Here we list all the diffusion algebras with $N=4$ generators which
can be useful for stochastic processes with
four species of particles (one can choose all the rates non-negative).

We first describe those of the algebras from the seven
series described in Sec.4
(see Eqs.(\ref{a1a})--(\ref{d1b})).
For the reader's convenience we point out all the $N=3$ subalgebras for each
algebra in the list.
The reader may be surprised by the fact that in the list which follows
the same symbol will denote two algebras
(see Eqs. (\ref{B3}) and (\ref{B5})), this
is due to the fact that as can be seen already in Eqs.(\ref{a1b})
and (\ref{a6a}) the
same notation is used for several algebras.

\begin{enumerate}
\item[\large {\bf 1.}~]
$A_I(4)$.
\be
\lb{B1}
g\ [ \D_i ,\ \D_j ] \ =  x_j\ \D_i - x_i\ \D_j\ , \qquad i, j = 0,1,2,3\ .
\ee
All the $N=3$ subalgebras of this algebra (i.e., the subalgebras generated by
different triples of the generators $\D_i$, $i=0,1,2,3$)
are of type $A_I$ (\ref{a5}).
\vspace{3mm}
\hrule

\item[\large {\bf 2.}~]
$A_I^{(1)}(3,1)$.
\be
\lb{B2}
\left\{
\begin{array}{l}
g\ [ \D_i ,\ \D_j ] \ =  x_j\ \D_i - x_i\ \D_j\ , \qquad i, j = 1,2,3\ ,
\\[2mm]
g_o\ [ \D_0 ,\ \D_i ] \ =  x_i\ \D_0\ .
\end{array}
\right.
\ee
This algebra contains the $A_I$ subalgebra $\{\D_1,\D_2,\D_3\}$
and three $B^{(1)}$ subalgebras (\ref{b3})
in which one takes $\Lambda=0$ and
which are
generated by the triples $\{\D_0,\D_1,\D_2\}$,
$\{\D_0,\D_1,\D_3\}$ and $\{\D_0,\D_2,\D_3\}$.
\vspace{3mm}
\hrule

\item[\large {\bf 3.}~]
$A_I^{(2)}(3,1)$ (two algebras).
\be
\lb{B3}
\phantom{a}\hspace{-10mm}
\begin{array}{ll}
\left\{
\begin{array}{l}
g\ [ \D_i ,\ \D_j ] \ =  x_j\ \D_i - x_i\ \D_j\ ,
\\[2mm]
g_o\  \D_0\ \D_i  \ =  x_i\ \D_0\ ,
\end{array}
\right.
&
\phantom{a}\hspace{-5mm}
\mbox{and~}
\left\{
\begin{array}{l}
g\ [ \D_i ,\ \D_j ] \ =  x_j\ \D_i - x_i\ \D_j\ ,
\\[1mm]
g_+\ \D_1\ \D_0  \ =  - x_1\ \D_0\ ,
\\[1mm]
g_-\  \D_0\ \D_{2,3} \ =  x_{2,3}\ \D_0\ ,
\end{array}
\right.
\\[3mm]
\phantom{a}\quad i, j = 1, 2, 3. &\qquad\;\;\quad i, j = 1, 2, 3.
\end{array}
\ee
Both algebras presented in (\ref{B3}) contain  the $A_I$
triple $\{\D_1,\D_2,\D_3\}$ and the $B^{(4)}$ triple
$\{\D_0,\D_2,\D_3\}$ (see (\ref{b8}) specialized to $\Lambda=0$).
The triples  $\{\D_0,\D_1,$ $\D_2\}$ and $\{\D_0,\D_1,\D_3\}$
are of $B^{(4)}$ type for the first case in (\ref{B3})
and they belong to the type $B^{(2)}$ for the last case
(see (\ref{b4}) with the special choice
$g_{\alpha\gamma}=g_{\gamma\alpha}=g$).

There are two more algebras of type $A_I^{(2)}(3,1)$.
Their defining relations are obtained by inverting the order of the
generators in all the products in formulae (\ref{B3}). This means that the
last pair of algebras would describe physical processes which are just
mirror reflections of the processes corresponding to the algebras
(\ref{B3}).
\vspace{3mm}
\hrule

\item[\large {\bf 4.}~]
$A_{II}(4)$.
\be
\lb{B4}
g_{ij}\  \D_i\  \D_j  \ =  x_j\ \D_i - x_i\ \D_j\ , \quad
\mbox{where}\quad i, j = 0,1,2,3\ , \;\; \mbox{and}\;\; i<j\ .
\ee
Here $g_{ij}:=g_i-g_j$ and $g_0>g_1>g_2>g_3$ so that $g_{ij}>0$ for all
$i<j$.

The algebra (\ref{B4}) contains only $A_{II}$ subalgebras (\ref{a6}).
\vspace{3mm}
\hrule

\item[\large {\bf 5.}~]
$A_{II}(3,1)$ (two algebras).
\be
\lb{B5}
\phantom{a}\hspace{-10mm}
\begin{array}{lcl}
\left\{
\begin{array}{l}
g_{ij}\  \D_i\ \D_j  \ =  x_j\ \D_i - x_i\ \D_j\ ,
\\[2mm]
g_{0i}\  \D_0\ \D_i  \ =  x_i\ \D_0\ ,
\end{array}
\right.
&\mbox{and}&
\left\{
\begin{array}{l}
g_{ij}\  \D_i\ \D_j  \ =  x_j\ \D_i - x_i\ \D_j\ ,
\\[1mm]
g_+\ \D_1\ \D_0  \ =  - x_1\ \D_0\ ,
\\[1mm]
g_{-2}\  \D_0\ \D_2 \ =  x_2\ \D_0\ ,
\\[1mm]
g_{-3}\  \D_0\ \D_3 \ =  x_3\ \D_0\ .
\end{array}
\right.
\end{array}
\ee
Here  both $i$ and $j$ take the values 1, 2, or 3; furthermore  one has
$i<j$ in the first relations in (\ref{B5});
$g_{ij}:=g_i-g_j$, $g_{0i}:=g_0-g_i$
and $g_{-i}:=g_- - g_i$, where $g_0>g_1>g_2>g_3$ and $g_->g_2$.

The algebras in (\ref{B5}) contain  the $A_{II}$
triple $\{\D_1,\D_2,\D_3\}$ and the $B^{(4)}$ triple
$\{\D_0,\D_2,\D_3\}$ (see (\ref{b8}) with $g=\Lambda$).
The triples  $\{\D_0,\D_1,\D_2\}$ and $\{\D_0,\D_1,\D_3\}$
are of $B^{(4)}$ type for the first case in (\ref{B5})
and they belong to type $B^{(2)}$ for the last case
(see (\ref{b4}) with $g_{\gamma\alpha}=0$).

One can construct two more algebras which are mirror partners
of those listed in (\ref{B5}) (c.f. the case {\bf 3}).
\vspace{3mm}
\hrule

\item[\large {\bf 6.}~]
$B^{(1)}(2,2)$.
\be
\lb{B6}
\phantom{a}\hspace{16mm}
\left\{
\begin{array}{l}
g\  \D_0\ \D_3\ -(g-\Lambda)\ \D_3\ \D_0  =  x_3\ \D_0 - x_0\ \D_3\ ,
\\[1mm]
g_a\  \D_0\ \D_a\ -(g_a-\Lambda)\ \D_a\ \D_0  =
- x_0\ \D_a\ ,
\\[1mm]
g_a\  \D_a\ \D_3\ -(g_a-\Lambda)\ \D_3\ \D_a  =
x_3\ \D_a\ , \qquad a=1,2\ ,
\\[1mm]
g_{12}\ \D_1\ \D_2 - g_{21}\ \D_2\ \D_1\ = 0\ .
\end{array}
\right.
\ee
This algebra contains two $B^{(1)}$ triples (\ref{b3}):
$\{\D_0,\D_1,\D_3\}$ and $\{\D_0,\D_2,\D_3\}$,
and two $C^{(1)}$ triples (\ref{c1}):
$\{\D_0,\D_1,\D_2\}$ and $\{\D_1,\D_2,\D_3\}$.
\vspace{3mm}
\hrule

\item[\large {\bf 7.}~]
$B^{(2)}(2,\ 0,2,0)$.
\be
\lb{B7}
\phantom{a}\hspace{-6mm}
\left\{
\begin{array}{l}
g\  \D_0\ \D_3\ -(g-\Lambda)\ \D_3\ \D_0  =  x_3\ \D_0 - x_0\ \D_3\ ,
\\[1mm]
g_+\  \D_0\ \D_a  =  - x_0\ \D_a\ ,
\\[1mm]
g_-\ \D_a\ \D_3  =
x_3\ \D_a\ , \qquad\quad a=1,2\ ,
\\[1mm]
g_{12}\ \D_1\ \D_2 - g_{21}\ \D_2\ \D_1\ = 0\ .
\end{array}
\right.
\ee
This algebra contains two $B^{(2)}$ triples (\ref{b4}):
$\{\D_0,\D_1,\D_3\}$ and $\{\D_0,\D_2,\D_3\}$,
and two $C^{(1)}$ triples:
$\{\D_0,\D_1,\D_2\}$ and $\{\D_1,\D_2,\D_3\}$,
where in Eq.(\ref{c1}) one takes
$g_\beta=g_\gamma=\Lambda$, respectively $g_\beta=g_\gamma=0$.
\vspace{3mm}
\hrule

\item[\large {\bf 8.}~]
$B^{(2)}(2,\ 2,0,0)$.
\be
\lb{B8}
\phantom{a}\hspace{-6mm}
\left\{
\begin{array}{l}
g\  \D_2\ \D_3\ -(g-\Lambda)\ \D_3\ \D_2  =  x_3\ \D_2 - x_2\ \D_3\ ,
\\[1mm]
(h-\Lambda)\ \D_a\ \D_2  =  x_2\ \D_a\ ,
\\[1mm]
h\ \D_a\ \D_3  =
x_3\ \D_a\ , \qquad\quad\; a=0,1\ ,
\\[1mm]
g_{01}\ \D_0\ \D_1 - g_{10}\ \D_1\ \D_0\ = 0\ .
\end{array}
\right.
\ee
This algebra contains two $B^{(4)}$ triples (\ref{b8}):
$\{\D_0,\D_2,\D_3\}$ and $\{\D_1,\D_2,\D_3\}$,
and two $C^{(1)}$ triples (\ref{c1}) with
$g_\beta=g_\gamma=0$:
$\{\D_0,\D_1,\D_2\}$ and $\{\D_1,\D_2,\D_3\}$.

The algebra $B^{(2)}(2,\ 0,0,2)$ is a mirror partner of the
algebra above.
\vspace{3mm}
\hrule

\item[\large {\bf 9.}~]
$C(1,3)$.
\be
\lb{B9}
\left\{
\begin{array}{l}
g_a\  \D_0\ \D_a\ -(g_a-\Lambda)\ \D_a\ \D_0  =  - x_0\ \D_a\ ,
\\[2mm]
g_{ab}\ \D_a\ \D_b - g_{ba}\ \D_b\ \D_a\ = 0\ , \qquad
a,b = 1,2,3\ .
\end{array}
\right.
\ee
This algebra contains three $C^{(1)}$ triples
$\{\D_0,\D_1,\D_2\}$, $\{\D_0,\D_1,\D_3\}$ and $\{\D_0,$ $\D_2,\D_2\}$,
and the triple $\{\D_1,\D_2,\D_3\}$ of the type $D$ (\ref{d}).

For $\Lambda\neq 0$, one can do the shift $\D_0\rightarrow \D_0-x_0/\Lambda$
and bring this algebra to the subcase of the family of quantum hyperplanes
(\ref{B10}).
\vspace{3mm}
\hrule

\item[\large {\bf 10.}~]
$D(0,4)$ algebra, or quantum hyperplane.
\be
\lb{B10}
g_{ab}\ \D_a\ \D_b - g_{ba}\ \D_b\ \D_a\ = 0\ , \qquad
a,b = 0,1,2,3\ .
\ee
Obviously, all the triples here are of $D$ type.
\vspace{3mm}
\hrule
\end{enumerate}

Next, we use the procedure of blending several diffusion algebras
as described in the Theorem at the end of Sec.4.
There is no need to point out anymore the $N=3$ subalgebras
for each example separately, since
the pair of main $N=3$ constituents
which are blended to produce an $N=4$ algebra
are mentioned explicitly in each case.
The remaining two $N=3$ subalgebras always belong to the
type $C^{(2)}$  (\ref{c2}).

Note that blending algebras of the types $A_I$ and $A_{II}$
produces only examples with $N\geq 5$.

One can get $N=4$ diffusion algebras by blending
any two of the following $N=3$ type $B$ algebras:
$B^{(1)}(2,1)$ ($\equiv B^{(1)}$ in the notations of Sec.3),
$B^{(2)}(2,\ 1,0,0)$ ($\equiv B^{(4)}$),
$B^{(2)}(2,\ 0,1,0)$ ($\equiv B^{(2)}$), and
$B^{(2)}(2,\ 0,0,1)$ ($\equiv B^{(3)}$).
The results are listed below.

\begin{enumerate}
\item[\large {\bf 11.}~]
Gluing $B^{(1)}$ (with generators $\{\D_0, \D_1, \D_3\}$) and
$B^{(2)}$
(with generators $\{\D_0, \D_2,$ $ \D_3\}$)
one obtains
\be
\lb{B11}
\left\{
\begin{array}{l}
g\  \D_0\ \D_3\ -(g-\Lambda)\ \D_3\ \D_0  =  x_3\ \D_0 - x_0\ \D_3\ ,
\\[1mm]
g_1\  \D_0\ \D_1\ -(g_1-\Lambda)\ \D_1\ \D_0  =
- x_0\ \D_1\ ,
\\[1mm]
g_1\  \D_1\ \D_3\ -(g_1-\Lambda)\ \D_3\ \D_1  =
x_3\ \D_1\ ,
\\[1mm]
g_+\ \D_0\ \D_2  = - x_0\ \D_2\ . \qquad
g_-\ \D_2\ \D_3  =
x_3\ \D_2\ ,
\end{array}
\right.
\ee
This set of relations should be supplemented by the condition
\be
\lb{B11a}
g_{12}\ \D_1\ \D_2\ = 0\ .
\ee
%
The algebra (\ref{B11}), (\ref{B11a}) has a mirror partner with an opposite order of indices 1 and 2 (one uses the condition $g_{21}\D_2 \D_1 =0$
instead of (\ref{B11a}) for it).
\vspace{3mm}
\hrule

\item[\large {\bf 12.}~]
Gluing $B^{(2)}$ ( with generators \{$\D_0, \D_1, \D_3$\}) and
$B^{(2)}$
(with generators \{$\D_0, \D_2, \D_3$\})
one obtains
\be
\lb{B12}
\phantom{a}\hspace{-13mm}
\left\{
\begin{array}{l}
g\  \D_0\ \D_3\ -(g-\Lambda)\ \D_3\ \D_0  =  x_3\ \D_0 - x_0\ \D_3\ ,
\\[1mm]
 g_+\ \D_0\ \D_1  = - x_0\ \D_1\ , \qquad
g_-\ \D_1\ \D_3  =
x_3\ \D_1\ ,
\\[1mm]
 h_+\ \D_0\ \D_2  = - x_0\ \D_2\ , \qquad
h_-\ \D_2\ \D_3  =
x_3\ \D_2\ ,
\end{array}
\right.
\ee
\be
\lb{B12a}
\mbox{and}\qquad g_{12}\ \D_1\ \D_2\ = 0\ .\hspace{65mm}
\ee
\hrule

\item[\large {\bf 13.}~]
Gluing $B^{(1)}$ (with generators \{$\D_1, \D_2, \D_3$\}) and
$B^{(4)}$
(with generators \{$\D_0, \D_2, \D_3$\})
one obtains
\be
\lb{B13}
\phantom{a}\hspace{-10mm}
\left\{
\begin{array}{l}
g\  \D_2\ \D_3\ -(g-\Lambda)\ \D_3\ \D_2  =  x_3\ \D_2 - x_2\ \D_3\ ,
\\[1mm]
(h-\Lambda)\ \D_0\ \D_2  =  x_2\ \D_0\ , \qquad
h\ \D_0\ \D_3  =  x_3\ \D_0\ ,
\\[1mm]
(g_1-\Lambda)\ \D_1\ \D_2\ - g_1\ \D_2\ \D_1 =  x_2\ \D_1\ ,
\\[1mm]
g_1\ \D_1\ \D_3\ - (g_1-\Lambda)\ \D_3\ \D_1 =  x_3\ \D_1\ ,
\end{array}
\right.
\ee
\be
\lb{B13a}
\mbox{and either}\quad g_{01}\ \D_0\ \D_1\ = 0\ , \quad
\mbox{or}\quad g_{10}\ \D_1\ \D_0\ =0\ . \hspace{34mm}
\ee
Gluing the algebras $B^{(1)}$ and $B^{(3)}$ produces a mirror partner of this algebra.
\vspace{3mm}
\hrule

\item[\large {\bf 14.}~]
Gluing $B^{(4)}$ (with generators \{$\D_1, \D_2, \D_3$\}) and
$B^{(4)}$
(with generators \{$\D_0, \D_2, \D_3$\})
one obtains
\be
\lb{B14}
\phantom{a}\hspace{1mm}
\left\{
\begin{array}{l}
g\  \D_2\ \D_3\ -(g-\Lambda)\ \D_3\ \D_2  =  x_3\ \D_2 - x_2\ \D_3\ ,
\\[1mm]
(h-\Lambda)\ \D_0\ \D_2  =  x_2\ \D_0\ , \qquad
h\ \D_0\ \D_3  =  x_3\ \D_0\ ,
\\[1mm]
(f-\Lambda)\ \D_1\ \D_2  =  x_2\ \D_1\ , \qquad
f\ \D_1\ \D_3  =  x_3\ \D_1\ ,
\end{array}
\right.
\ee
\be
\lb{B14a}
\mbox{and}\qquad g_{01}\ \D_0\ \D_1\ = 0\ . \hspace{67mm}
\ee
Gluing the algebras $B^{(3)}$ and $B^{(3)}$ produces a
mirror partner of this algebra.
\vspace{3mm}
\hrule

\item[\large {\bf 15.}~]
Gluing $B^{(3)}$ (with generators \{$\D_1, \D_2, \D_3$\}) and
$B^{(4)}$
(with generators \{$\D_0, \D_1, \D_2$\})
one obtains
\be
\lb{B15}
\phantom{a}\hspace{7mm}
\left\{
\begin{array}{l}
g\  \D_1\ \D_2\ -(g-\Lambda)\ \D_2\ \D_1  =  x_2\ \D_1 - x_1\ \D_2\ ,
\\[1mm]
h\ \D_1\ \D_3  = - x_1\ \D_3\ , \qquad
(h-\Lambda)\ \D_2\ \D_3  =  - x_2\ \D_3\ ,
\\[1mm]
(f-\Lambda)\ \D_0\ \D_1  =  x_1\ \D_0\ , \qquad
f\ \D_0\ \D_2  =  x_2\ \D_0\ ,
\end{array}
\right.
\ee
\be
\lb{B15a}
\mbox{and}\qquad g_{03}\ \D_0\ \D_3\ = 0\ . \hspace{67mm}
\ee
\hrule

\item[\large {\bf 16.}~]
Gluing $B^{(2)}$ (with generators \{$\D_1, \D_2, \D_3$\}) and
$B^{(4)}$
(with generators \{$\D_0, \D_1, \D_3$\})
one obtains
\be
\lb{B16}
\phantom{a}\hspace{-13mm}
\left\{
\begin{array}{l}
g\  \D_1\ \D_3\ -(g-\Lambda)\ \D_3\ \D_1  =  x_3\ \D_1 - x_1\ \D_3\ ,
\\[1mm]
g_+\ \D_1\ \D_2  = - x_1\ \D_2\ , \qquad
g_-\ \D_2\ \D_3  =   x_3\ \D_2\ ,
\\[1mm]
(h-\Lambda)\ \D_0\ \D_1  =  x_1\ \D_0\ , \qquad
h\ \D_0\ \D_3  =  x_3\ \D_0\ ,
\end{array}
\right.
\ee
\be
\lb{B16a}
\mbox{and}\qquad g_{02}\ \D_0\ \D_2\ = 0\ . \hspace{67mm}
\ee
Gluing the algebras $B^{(2)}$ and $B^{(3)}$ produces mirror partner of this algebra.
\vspace{3mm}
\hrule
\end{enumerate}
Gluing a pair of $B^{(1)}$ algebras gives a special case
of the $B^{(1)}(2,2)$ algebra (see case 6 above) with $g_{21}=0$.

There are two other possibilities to blend the $C$ type algebras
$C(1,2)$ and $C(1,1)$ (for their definition see Eq.(\ref{c1b}))
into a $N=4$ diffusion algebra.

\begin{enumerate}

\item[\large {\bf 17.}~]
Gluing $C(1,2)$ (with generators \{$\D_0, \D_1, \D_2$\}) and
$C(1,1)$
(with generators \{$\D_0, \D_3$\})
one obtains
\be
\lb{B17}
\phantom{a}\hspace{10mm}
\left\{
\begin{array}{l}
g_a\  \D_0\ \D_a\ -(g_a-\Lambda)\ \D_a\ \D_0  =  - x_0\ \D_a\ , \qquad
a=1,2\ ,
\\[1mm]
g_{12}\ \D_1\ \D_2\ - g_{21}\ \D_2\ \D_1\ = 0\ ,
\\[1mm]
g_{03}\  \D_0\ \D_3\ -g_{30}\ \D_3\ \D_0  =  - x_0\ \D_3\ ,
\end{array}
\right.
\ee
with either one of the following two sets of conditions:
\be
\lb{B17a}
g_{13}\ \D_1\D_3=g_{23}\ \D_2\D_3 = 0\ , \quad\mbox{or}\quad
g_{13}\ \D_1\D_3=g_{32}\ \D_3\D_2 =0\ .
\ee
In this algebra, besides the $C^{(1)}$ triple $\{\D_0,\D_1,\D_2\}$
there are two $C^{(2)}$ triples $\{\D_0,\D_1,\D_3\}$ and
$\{\D_0,\D_2,\D_3\}$ and the $D$ triple $\{\D_1,\D_2,\D_3\}$.
\vspace{3mm}
\hrule

\item[\large {\bf 18.}~]
Gluing three  copies of
$C(1,1)$ algebra
one obtains
\be
\lb{B18}
\begin{array}{l}
g_{0a}\  \D_0\ \D_a\ -g_{a0}\ \D_a\ \D_0  =  - x_0\ \D_a\ , \qquad
a=1,2,3\ ,
\end{array}
\ee
\be
\lb{B18a}
\mbox{and}\qquad g_{12}\ \D_1\D_2=g_{13}\ \D_1\D_3=g_{23}\ \D_2\D_3 = 0\ .
\hspace{40mm}
\ee
This algebra contains one $D$ type triple $\{\D_1,\D_2,\D_3\}$
and all the other triples are of the type $C^{(2)}$.
\vspace{3mm}
\end{enumerate}



\begin{thebibliography}{99}

\bibitem{D}T.M.Liggett, "Interacting Particle Systems" (Springer, New York, 1985); \\
T.M.Liggett, Ann.Prob. {\bf 25} (1997) 1; \\
B.Derrida, Physics Reports {\bf 301} (1998) 65 and references therein;\\
M.R.Evans, {\tt cond-math/0007293} and references therein; \\
G.M.Schuetz, in "Phase Transitions and Critical Phenomena",
Eds. C.Domb and J.L.Lebowitz (Academic Press, London, 2000).

\bibitem{E}M.R.Evans, Europhys. Lett. {\bf 36} (1996) 13.

\bibitem{SD} G.Schuetz and E.Domany, J. Stat. Phys. {\bf 72} (1993) 277.

\bibitem{ARR} P.F.Arndt, V.Rittenberg and S.Radic, to be published.

\bibitem{DEHP} B.Derrida, M.R.Evans, V.Hakim and V.Pasquier, J. Phys. A: Math. Gen. {\bf 26} (1993) 1493.

\bibitem{S} G.M.Schuetz,  in {\em "Statistical Models, Yang-Baxter Equations
and related topics"}. Eds. M.L.Ge and F.Y.Wu, World Scientific, Singapore, 1996.

\bibitem{DJLS} B.Derrida, S.A.Janowsky, J.L.Lebowitz and E.R.Speer,
J.Stat.Phys. {\bf 73} (1993) 813.

\bibitem{IPR} A.P.Isaev, P.N.Pyatov and V.Rittenberg, to be published.

\bibitem{ADR} F.C.Alcaraz, S.Dasmahapatra and V.Rittenberg, J. Phys. A: Math. Gen.
{\bf 31} (1998) 845 and references therein.

\bibitem{KS}  K.Krebs and S.Sandow, J. Phys. A: Math. Gen. {\bf 30} (1997) 3165.

\bibitem{ER} F.H.L.Essler and V.Rittenberg, J. Phys. A: Math. Gen. {\bf 29}
(1996) 3375; \\
K.Mallick and S.Sandow,
J. Phys. A: Math. Gen. {\bf 30} (1997) 4513-4526; \\
R.A.Blythe, M.R.Evans, F.Colaiori and F.H.L.Essler, J.Phys. A:
Math. Gen. {\bf 33} (2000) 2313.


\bibitem{AHR} P.F.Arndt, T.Heinzel and V.Rittenberg, J. Phys. A: Math. Gen.
{\bf 31} (1998) 833 and references therein.

\bibitem{K} V.Karimipour, Phys. Rev. E {\bf 59} (1999) 205.


\bibitem{Bergman} Bergman, G.M.:
{\em `The diamond lemma for ring theory'.}
Adv. in Math. {\bf 29} (1978) 178-218.

\bibitem{PBW}H.Poincar\'{e}, {\em "Sur les groupes continus"}, Trans. Cambridge
Philos. Soc. {\bf 18} (1900) 220;\\
G.Birkhoff, Ann. of Math. {\bf 38} (1937) 526;\\
E. Witt, {\em "Treue Darstellung Liescher Ringe"}, J. Reine Angew. Math.
{\bf 117} (1937) 152.

\bibitem{M} Yu.I.Manin,
{\em "Quantum groups and Noncommutative Geometry"},
Montreal University Preprint CRM-1561 (1989); \\
Yu.I.Manin, Comm. Math. Phys. {\bf 122} (1989) 163; \\
A.Sudbery, J. Phys. A: Math. Gen. {\bf 23} (1990) L697.

\bibitem{DJ} N.Reshetikhin, Lett. Math. Phys. {\bf 20} (1990) 331;\\
A.P.Isaev, Phys. Part. Nucl. {\bf 26}, No.5 (1995) 501.

\bibitem{BIO} D.Bernard, Phys. Lett. {\bf B 260} (1991) 389; \\
C.Burdik, A.P.Isaev and O.V.Ogievetsky,
{\em "Standard complex for quantum Lie algebras"}, (2000), math.QA/0010060; \\
A. Sudbery, {\em "Quantum Lie algebras of type $A(N)$"}, (1995), q-alg/9510004.

\bibitem{CG}E.Cremmer and J.-L.Gervais, Comm. Math. Phys. {\bf 134} (1990) 619.

\bibitem{AR} F.C.Alcaraz and V.Rittenberg, J. Phys. A: Math. Gen {\bf 33}
(2000) 7469 and references therein.



\end{thebibliography}
\end{document}